\documentclass[sigconf,nonacm]{acmart}

\usepackage{amsmath}
\usepackage{booktabs}
\usepackage{tabularx}
\usepackage{multirow}
\usepackage{stfloats}
\setcopyright{none}
\newcommand{\method}{PACT}

\newcommand{\pool}{\mathcal{P}}
\newcommand{\sources}{\mathcal{S}}
\newcommand{\E}{\mathbb{E}}

\newcommand{\loss}{\mathcal{L}}

\title[PACT: Defending Recommendation against Promotional News Rewriting]{PACT: Defending Recommendation against Promotional News Rewriting via Attack-Diverse Score Consistency}
\author{Chongxian Chen}
\affiliation{\institution{Waseda University}\city{Tokyo}\country{Japan}}
\author{Fan Mo}
\affiliation{\institution{ZOZO NEXT, Inc.}\city{Chiba}\country{Japan}}
\affiliation{\institution{Waseda University}\city{Tokyo}\country{Japan}}
\author{Xin Fan}
\affiliation{\institution{Waseda University}\city{Tokyo}\country{Japan}}
\author{Hayato Yamana}
\affiliation{\institution{Waseda University}\city{Tokyo}\country{Japan}}

\begin{document}
\raggedbottom
\begin{abstract}
\begingroup
\emergencystretch=1em
Personalized news recommenders are increasingly vulnerable to LLM-powered promotional attacks, where content providers rephrase headlines to manipulate rankings without altering underlying facts. Defending against such manipulation requires covering a wide spectrum of LLM generators and optimization strategies---ranging from stylistic prompting to reinforcement feedback---without incurring prohibitive training or inference costs. In this work, we present \textbf{\method{}} (\textbf{P}ooled \textbf{A}ttack \textbf{C}onsistency \textbf{T}raining), a model-agnostic defense that decouples relevance estimation from superficial rhetorical variations. PACT integrates three synergistic operations: it \textbf{pools} offline, fact-checked rewrites from multi-generator and multi-strategy attack sources; \textbf{samples} a constant number of variants at each training visit to yield an unbiased estimator of the pool-wide consistency objective under a fixed compute budget; and \textbf{aligns} original and rewritten ranking scores within the same user context jointly with click supervision. Requiring no architectural modifications or test-time filtering, PACT deploys the trained recommender with zero extra latency. Across nine experimental settings spanning three recommendation backbones and three unseen-generator attacks, PACT consistently achieves the lowest promotion rate and target exposure, outperforming the strongest defense baseline by up to 3.0 percentage points in promotion reduction under LANCE. Further evaluations verify that PACT's superiority holds robustly under unseen attack-strategy holdouts and adaptive white-box retraining, while keeping clean recommendation utility virtually intact.
\par
\endgroup
\end{abstract}

\ccsdesc[500]{Information systems~Recommender systems}
\ccsdesc[300]{Computing methodologies~Natural language generation}
\keywords{News recommendation, promotional rewriting, adversarial robustness, attack diversity, consistency learning}
\maketitle

\section{Introduction}
\label{sec:introduction}

Personalized news recommendation matches article content to users' reading histories, helping readers navigate a rapidly changing information supply. Neural recommenders learn this matching through attentive text encoders and user representations~\citep{naml2019,nrms2019,lstur2019}, while pretrained language models further improve the modeling of news semantics~\citep{wu2021plmnr}. These models depend on how news is expressed to estimate its relevance. The same dependence, however, gives content providers an opportunity to influence recommendation scores by changing the wording of a headline.

\begin{figure}[t]
\centering
\includegraphics[width=\columnwidth]{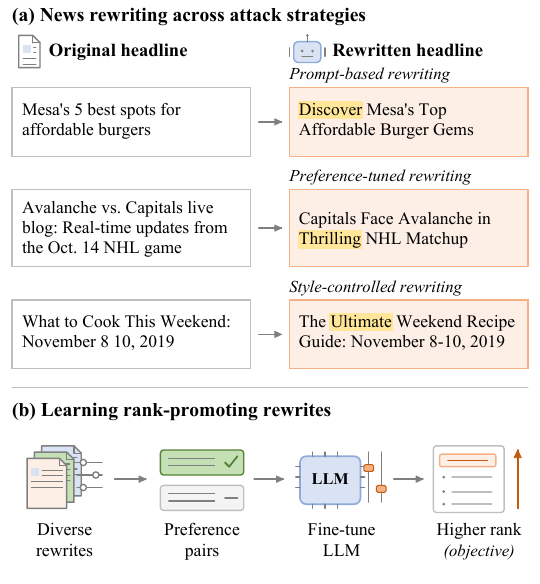}
\caption{Promotional news rewriting. (a) Actual headline rewrites from prompt-based, preference-tuned, and style-controlled rewriting; yellow highlights rhetorical cues. (b) Diverse rewrites and preference pairs are used to fine-tune an LLM toward rank promotion~\citep{lance2025}. The ranked list illustrates the attack objective.}
\Description{Three original MIND headlines are paired with actual outputs from prompt-based, preference-tuned, and style-controlled rewriting. Discover, Thrilling, and Ultimate are highlighted. A schematic connects diverse rewrites, preference pairs, LLM fine-tuning, and the objective of promoting a target news item in a ranked list. News, parameter-adjustment, and ranking icons accompany the flow.}
\label{fig:motivation}
\end{figure}

Large language models (LLMs) make such manipulation practical at scale. A provider can vary a headline's tone, emphasis, or rhetorical style while retaining its underlying facts, as illustrated in Figure~\ref{fig:motivation}(a). Prior studies demonstrate that textual changes can promote items in LLM-based and ID-free recommenders~\citep{textrecattack2024,textsimu2025}; ATR and LANCE further develop rewriting attacks that optimize for ranking gains in text-aware and news recommendation~\citep{atr2024,lance2025}, as illustrated in Figure~\ref{fig:motivation}(b). A fluent, fact-preserving rewrite can therefore seek additional exposure without introducing a false claim. Related source-bias studies find preferences for generated text in retrieval~\citep{retrieverbias2024} and their amplification in simulated recommendation feedback loops~\citep{sourcebias2025}. These findings motivate defenses against expression-induced promotion that also retain useful personalization.

Existing work provides useful foundations for this defense. Score alignment mitigates source preference in recommendation, including across generators and in environments containing text from multiple LLMs~\citep{sourcebias2025}. SheepDog combines diverse reframing with prediction consistency to improve style robustness in fake-news detection~\citep{sheepdog2024}. These results support learning across alternative expressions, but leave open how to cover rewrites deliberately optimized for personalized ranking. Such attacks vary along two dimensions: the generator and the rewriting strategy. In particular, recommendation feedback can select expressions beyond prescribed stylistic changes~\citep{atr2024,lance2025}. Covering these variations also raises a training-budget question, since evaluating every available rewrite at each update makes auxiliary computation grow with the pool. The resulting challenge is to expose a recommender to diverse promotional rewrites within a fixed per-update budget while preserving relevance learning.

Our key idea is to \emph{vary the expression while stabilizing the recommendation score}. For screened rewrites assumed to preserve relevance-bearing content, we regularize the score of the same news under the same user context. Writing style can itself carry information about user preferences~\citep{esm2023}. We therefore combine score consistency on screened pairs with click supervision for relevance learning. Motivated by augmentation-based consistency learning~\citep{uda2020} and exposure to multiple attack sources~\citep{ensemble_at2018}, we vary both generators and rewriting strategies, and distribute exposure to their outputs over successive training visits.

We instantiate this idea as \method{} (Pooled Attack Consistency Training), with three operations: \textbf{pool, sample, and align}. First, we pool screened, deduplicated rewrites from multiple generators and prompted or feedback-optimized attack strategies around each news identity. Second, we sample a fixed number of variants at each eligible candidate occurrence, refreshing the sampled expressions across training visits. Third, we align original and rewritten scores under the same user context alongside the ordinary click loss. Sampling provides an unbiased estimate of the pool-averaged consistency objective while keeping the number of auxiliary views per occurrence independent of pool size. Rewrites are generated offline.

We evaluate \method{} on MIND-small~\citep{mind2020} with three news recommendation backbones, testing transfer to a held-out generator family and unseen attack strategies, recommendation utility under LANCE attack, pool composition, and attacks retrained against the defended models. \method{} achieves lower promotion frequency and target exposure than the defense baselines across all nine main attack--backbone combinations. Its advantage in rank gain and promotion frequency persists under strategy holdouts and retrained attacks. Recommendation utility under LANCE exhibits a backbone-dependent trade-off: mean relative changes across the reported metrics range from $-1.26\%$ to $+0.74\%$.

Our contributions are threefold:
\begin{enumerate}
  \item \textbf{Attack-diverse training pools.} We organize screened promotional rewrites by news identity, combining generator and attack-strategy diversity to cover both prompted and feedback-optimized manipulation.
  \item \textbf{\method: pool, sample, and align.} We develop a defense that samples from these pools to train score consistency under a fixed auxiliary view budget, alongside click supervision.
  \item \textbf{Robustness and utility evaluation.} We assess transfer to unseen generators and strategies, resistance to retrained attacks, and recommendation utility, with controlled comparisons of pool composition.
\end{enumerate}

\section{Related Work}
\label{sec:related}

\noindent\textbf{News recommendation.}
Neural news recommenders combine content encoding with interest modeling. Representative architectures incorporate knowledge entities, multiple content views, and attention mechanisms~\citep{wang2018dkn,naml2019,nrms2019,wu2019npa}. User representations capture long- and short-term preferences~\citep{lstur2019}, multiple interests~\citep{li2022miner}, and candidate-dependent relevance~\citep{qi2022caum}. Pretrained language models strengthen news representations~\citep{wu2021plmnr}, while LLM-enhanced methods enrich content and align semantic representations with collaborative signals~\citep{liu2024once,ren2024rlmrec}. S$^2$LENR further addresses structural information loss and semantic noise in LLM-based augmentation~\citep{s2lenr2025}. Beyond topical relevance, event--style matching models readers' preferences for both events and writing styles~\citep{esm2023}, highlighting the role of expression in personalized news selection.

\noindent\textbf{Textual attacks on recommendation.}
Recommendation poisoning attacks manipulate interaction sequences~\citep{loki2020} or perturb news content used for training~\citep{tdpcp2022}. In text classification, TextFooler and BERT-ATTACK search word substitutions under semantic constraints to alter model predictions~\citep{textfooler2020,bertattack2020}. Recent recommendation attacks directly modify item descriptions to increase exposure. TextRecAttack searches textual perturbations against LLM-based recommenders~\citep{textrecattack2024}, while TextSimu uses collaborating LLM agents to imitate characteristics of popular items~\citep{textsimu2025}. ATR explores fine-tuning and in-context learning for promotional rewriting~\citep{atr2024}. In news recommendation, LANCE explores styles, sentiments, and personas, then learns rewriting strategies from recommendation feedback~\citep{lance2025}. These approaches extend textual manipulation from local substitutions to generated rewrites, with varying requirements for model access and feedback.

\noindent\textbf{Content bias and robustness.}
Adversarial personalized ranking regularizes recommendation against parameter perturbations~\citep{apr2018}. In classification, UDA and AugMix combine augmentation with prediction consistency, including text augmentation in UDA and image mixtures in AugMix~\citep{uda2020,augmix2020}. TRADES studies the accuracy--robustness trade-off~\citep{trades2019}, and ensemble adversarial training incorporates attacks transferred from multiple models~\citep{ensemble_at2018}. Robustness evaluations also examine unforeseen perturbation types and attacks adapted to the defense~\citep{unforeseen2019,adaptive2020}.

For news text, cross-field contrastive learning reduces title bias by relating headlines to abstracts~\citep{titledebias2024}. Source-bias studies identify preferences for LLM-generated content in retrieval and recommendation~\citep{retrieverbias2024,sourcebias2025}. Proposed remedies include aligning recommendation scores across original and rewritten candidates and histories~\citep{sourcebias2025}, and aligning generators to reduce source bias in their outputs~\citep{llmsbm2025}. In fake-news detection, SheepDog combines style-diverse reframing, prediction consistency, and content-focused veracity supervision~\citep{sheepdog2024}. AdStyle iteratively generates and selects style-conversion prompts using detector confusion, semantic coherence, and diversity~\citep{adstyle2025}. These studies connect content variation to recommendation bias and classification robustness, providing the context for examining promotional rewriting in personalized ranking.

\section{Problem Formulation}
\label{sec:problem}

Let $n$ denote a news identity, $x_n=(t_n,c_n)$ its title and category, and $h_u$ the reading history associated with a user context $u$. A recommender with parameters $\theta$ assigns a score $f_\theta(h_u,x_n)$ and ranks eligible news by this score. We write $r_\theta(u,n;X)$ for the rank of item $n$ in a catalog whose news-feature state is $X$, with smaller values indicating better ranks. Both candidate and history representations are determined by $X$. Training examples consist of observed clicks and unclicked candidates within the same impression. The defense preserves this recommendation task and augments its training objective.

\paragraph{Promotional attacker.}
The attacker controls the headlines of a target set $\mathcal{T}$, replacing $x_n=(t_n,c_n)$ with $\widetilde{x}_n=(\widetilde{t}_n,c_n)$, where $\widetilde{t}_n$ is generated with instructions to remain fluent and preserve the original title's meaning. The objective is to improve target ranks and exposure across user contexts. Evaluation attack generation uses only the original title and category. The attacker does not change categories, news identities, click labels, user identities, or the candidate catalog. In the attacked evaluation world, the changed title is used wherever that news identity appears, including reading histories.

\paragraph{Attack source.}
A source $s=(g,a)$ consists of a generator $g$ and an attack strategy $a$; model versions, prompts, and decoding settings are recorded as source metadata. Let $\sources_{\mathrm{train}}$ contain the available training sources. The defender has original news, ordinary recommendation training data, and offline rewrites produced by these sources. At test time, the source may use an unseen generator, an unseen strategy, or both. Source coverage is a defense-training design variable.

\paragraph{Defense objective.}
We seek parameters that reduce rewriting-induced promotion while maintaining recommendation utility. For example, the average rank gain is
\begin{equation}
 G(\theta)=\frac{1}{|\mathcal{U}|\,|\mathcal{T}|}
 \sum_{u\in\mathcal{U}}\sum_{n\in\mathcal{T}}
 \big[r_\theta(u,n;X)-r_\theta(u,n;\widetilde{X})\big],
 \label{eq:rankgain}
\end{equation}
where $\mathcal{U}$ is the fixed set of evaluation contexts and $\widetilde{X}$ replaces all target texts simultaneously. We assess attack gains alongside recommendation AUC, MRR, and nDCG under LANCE attack. Score consistency provides the trainable surrogate; ranking and exposure metrics assess its effect on the recommendation outcome.

\section{\method: Diverse Exposure, Consistent Scores}
\label{sec:method}

\begin{figure*}[t]
\centering
\includegraphics[width=0.7\textwidth]{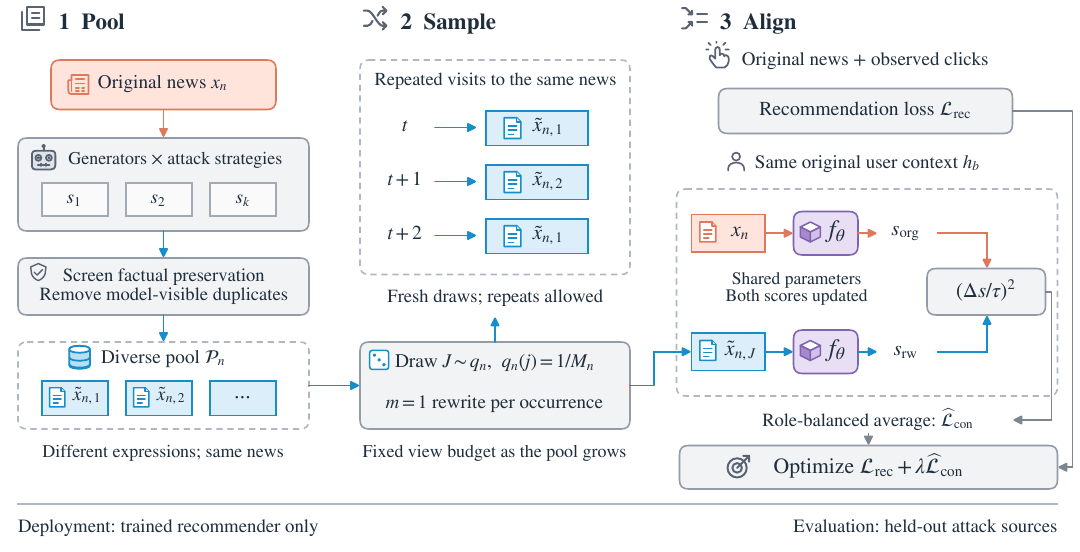}
\caption{\textbf{\method: vary the expression, stabilize the score.} (1) Pool screened rewrites from different generators and attack strategies. (2) Sample fresh variants at successive training visits, with replacement; $m=1$ is shown. (3) Align original and rewritten scores under the same user context alongside click supervision; $\Delta s=s_{\rm rw}-s_{\rm org}$.}
\Description{Three stages labeled Pool, Sample, and Align. Different sources produce screened rewrites of the same news. Independent sampling exposes the recommender to changing expressions across visits, with repeats allowed. Original and rewritten views share recommender parameters and user context. Score consistency and the original click loss jointly update trainable parameters. Deployment uses the trained recommender, and held-out attacks evaluate transfer.}
\label{fig:overview}
\end{figure*}

\method{} trains a recommender to preserve user-conditioned scores across screened promotional rewrites. Paired titles share a news identity and are assumed to preserve relevance-bearing content. Its guiding principle is to \emph{vary the expression, stabilize the score}. As shown in Figure~\ref{fig:overview}, it has three components: (1) \textbf{pool} rewrites from diverse generators and attack strategies around each news identity; (2) \textbf{sample} fresh variants whenever that news appears during training; and (3) \textbf{align} original and sampled scores while learning recommendation from observed clicks.

\subsection{Pool: Cover Different Routes to Promotion}
\label{sec:pool}

Rewriting diversity has two dimensions: \emph{who generates the text} and \emph{how the rewrite is obtained}. Prompted styles directly vary tone and framing; ATR-2FT learns domain-specific, rank-promoting rewrites~\citep{atr2024}; LANCE learns from exploration and recommendation-feedback preferences~\citep{lance2025}. Combining these routes exposes the defender to both prescribed stylistic changes and changes selected for their effect on recommendation.

For each eligible training news item $n$, we collect rewrites from available sources $s=(g,a)\in\sources_{\mathrm{train}}$, where $g$ is the generator and $a$ the attack strategy. We retain the news identity, category, and click labels, replacing only the title. Generation context varies by prompt family (Section~\ref{sec:pool_construction}). Screening checks the original title's entities, events, quantities, temporal qualifications, and attribution. We remove candidates that become identical to the original or another retained rewrite after the recommender's tokenization and truncation.

The result is a news-specific pool
\begin{equation}
 \pool_n=\{\widetilde{x}_{n,1},\ldots,\widetilde{x}_{n,M_n}\}.
\end{equation}
Pool sizes $M_n$ can differ. News with no admitted rewrite remain in ordinary recommendation training. We exclude test target IDs and any held-out sources before constructing the training pools. Generator and strategy provenance remain separate, so coverage can be expanded or held out along either dimension.

\subsection{Sample: Refresh Exposure at a Fixed Budget}
\label{sec:sampling}
\label{sec:cost}

For every eligible candidate occurrence $(b,n)$ in training impression $b$, we independently draw $m$ variants:
\begin{equation}
 J_{bn\ell}\sim q_n,\qquad q_n(j)=\frac{1}{M_n},
 \quad \ell=1,\ldots,m.
 \label{eq:sampling}
\end{equation}
The default is $m=1$. Draws are with replacement and independent across occurrences, including repeated appearances of the same news within a minibatch. Later visits can therefore present different expressions without enumerating the pool at every update. This operation samples rewrite identities; recommendation candidates and click labels remain unchanged.

Sampling fixes auxiliary work at $m$ views per eligible occurrence and gives each news occurrence equal weight regardless of pool size. Across visits, each variant receives selection probability $1/M_n$ per draw, distributing exposure over the available expressions. Candidate-uniform sampling matches the expected weights of the pool-averaged objective.

\subsection{Align: Learn Relevance across Expressions}
\label{sec:objective}

We train on observed clicks while penalizing score changes between an original title and its sampled rewrites. For minibatch $\mathcal{B}$, let $n_b^+$ be the clicked news and $\mathcal{C}_b$ contain it and the valid sampled negatives. The recommendation loss uses original text:
\begin{equation}
 \loss_{\mathrm{rec}}=-\frac{1}{|\mathcal{B}|}\sum_{b\in\mathcal{B}}
 \log\frac{\exp f_\theta(h_b,x_{n_b^+})}
 {\sum_{n\in\mathcal{C}_b}\exp f_\theta(h_b,x_n)}.
 \label{eq:rec}
\end{equation}
Padding is masked. Click supervision trains the model to distinguish relevant news.

Following score-alignment and logit-pairing objectives~\citep{sourcebias2025,alp2018}, we compare each pair under the same original user context $h_b$:
\begin{equation}
 d_\theta(b,n,j)=
 \left(\frac{f_\theta(h_b,\widetilde{x}_{n,j})-
 f_\theta(h_b,x_n)}{\tau}\right)^2.
 \label{eq:discrepancy}
\end{equation}
Here $\tau>0$ is a score scale estimated once from the clean starting checkpoint and then fixed. The symmetric penalty constrains score changes in both directions. Gradients propagate through both the original and rewritten score branches.

Let $\mathcal{E}_{\mathcal{B}}^+$ and $\mathcal{E}_{\mathcal{B}}^-$ denote positive and negative candidate occurrences with nonempty pools. We average within each role $r\in\{+,-\}$ and optimize
\begin{align}
 \widehat{D}_r & =
 \frac{1}{|\mathcal{E}_{\mathcal{B}}^r|}
 \sum_{(b,n)\in\mathcal{E}_{\mathcal{B}}^r}
 \frac{1}{m}\sum_{\ell=1}^{m}d_\theta(b,n,J_{bn\ell}),
 \label{eq:role}\\
 \widehat{\loss}_{\mathrm{con}}&=\tfrac12\widehat{D}_+
 +\tfrac12\widehat{D}_-,\qquad
 \loss=\loss_{\mathrm{rec}}+\lambda\widehat{\loss}_{\mathrm{con}}.
 \label{eq:joint}
\end{align}
An empty role contributes zero without rescaling the other. Equal role weights prevent the more numerous negatives from dominating through frequency alone. Paired auxiliary evaluations disable dropout to isolate the text change; the recommendation branch retains its usual training dropout. We update the recommender's trainable parameters from a clean checkpoint, retaining its existing frozen parameters. Pool generation is offline. Appendix~\ref{app:defense_details} specifies scale calibration and implementation details.

\paragraph{Sampling objective and transfer.}
For a fixed batch and model state, Equation~\ref{eq:joint} is an unbiased estimator of the pool-averaged objective (Appendix~\ref{app:method}). Training therefore distributes consistency constraints across the stored rewriting mechanisms at a fixed view budget. When attack sources exploit shared expression cues, these constraints may also suppress gains from held-out sources. We evaluate this transfer through ranking and exposure metrics. At inference, \method{} uses the trained recommender directly.

\section{Experimental Setup}
\label{sec:exp}

We evaluate five research questions: robustness to promotional rewriting from a held-out model family (RQ1), generalization to unseen attack strategies (RQ2), the effect of defense on recommendation performance under LANCE attack (RQ3), the effect of pool composition (RQ4), and robustness to attacks retrained against the evaluated recommender (RQ5).

\subsection{Dataset and Evaluation Protocol}
\label{sec:data_backbones}

We use MIND-small~\citep{mind2020}, which contains 65,238 news articles and 94,057 users across the official training and development files. We split the official training impressions chronologically: November 9--13, 2019 is used for training and November 14 for validation. The official development set is used for testing. The resulting splits contain 126,695, 30,270, and 73,152 impressions, respectively. Model selection uses only the validation split. Each clicked candidate forms one training instance, yielding 189,519 instances. We sample up to four distinct unclicked candidates from the same impression for each positive; missing negatives are padded and masked. Recommendation evaluation retains the complete candidate list of each impression.

Following LANCE~\citep{lance2025}, we use 3,000 exploration news items for attack training and 300 target news items for attack evaluation. Exploration and target news are sampled from the training and test news, respectively. Each set covers three popularity strata: the 0--20th, 40--60th, and 80--100th frequency percentiles, with 1,000 exploration items and 100 targets per stratum. The target set is held fixed across all attacks, defenses, and recommendation models. Attack-training news and defense-pool news exclude test target IDs.

All 300 target titles are replaced simultaneously, both in candidate news and wherever those articles appear in a user's history. Full-catalog ranking uses the 42,416 real articles in the official development catalog and 50,000 distinct user--history pairs. Original and attacked evaluations use the same checkpoint, catalog, user contexts, and target IDs. Each target receives one generated rewrite, without test-score selection or retries. Structurally invalid generations fall back to the original title and remain in the denominator. Main-table evaluation uses the generated outputs without semantic filtering. Test data are not used to select hyperparameters, checkpoints, or generated candidates.

\subsection{Recommendation Backbones}
\label{sec:backbones}

Following LANCE~\citep{lance2025}, we evaluate three established news recommendation models: NAML~\citep{naml2019}, NRMS~\citep{nrms2019}, and LSTUR~\citep{lstur2019}. All models use news titles and categories together with users' clicked-news histories. Titles are truncated to 30 tokens, and histories retain at most 50 articles. Common preprocessing and training settings are provided in Appendix~\ref{app:recommender_details}.

\subsection{Attack Methods and Defense Baselines}
\label{sec:attack_setup}
\label{sec:baselines}

We compare prompt-based rewriting, ATR-2FT, and LANCE. For these evaluation attacks, all LLMs receive the original news title and category, and generate a rewritten title. Rewriting changes only the title; categories, click labels, and article identities remain fixed. We use GPT-4o as the LANCE Explorer, Gemma 4 31B for prompt-based rewriting, and Gemma 4 E4B as the trainable base model for ATR-2FT and the LANCE Reflector. In RQ1--RQ4, ATR-2FT and LANCE are trained against the undefended recommender, and each attack's generated titles are held fixed across defenses. RQ5 instead retrains these attacks against each evaluated checkpoint. Appendix~\ref{app:attack_details} provides training and decoding settings.

\paragraph{Prompt-based rewriting.}
We prompt Gemma 4 31B to rewrite each target headline for engagement and recommendation relevance while preserving its original meaning. The prompt contains the original title and category. This baseline uses the same headline-rewriting instruction and decoding settings as the LANCE Reflector, without attack fine-tuning or recommender feedback.

\paragraph{ATR-2FT w/ Gemma 4 E4B}
ATR-2FT~\citep{atr2024} first adapts the language model to the news domain and then jointly optimizes text generation and rank promotion. We follow the news adaptation described by LANCE~\citep{lance2025}: extract the frozen recommender's news encoder, align the language-model representations with its news-embedding space, and use the aligned representations to compute the promotion loss. Language-model updates receive gradients from the promotion and generation objectives, while the recommender remains frozen. Attack training uses exploration news, and generation is evaluated on the separate 300-target set. Appendix~\ref{app:atr_details} defines the objectives.

\paragraph{LANCE w/ Gemma 4 E4B}
LANCE~\citep{lance2025} consists of an Explorer and a Reflector. The Explorer uses GPT-4o to generate headline variants across five writing styles, three sentiments, and six personas, conditioned on the original title and category. Recommendation feedback is used to construct preferred and dispreferred rewrite pairs. The Reflector is initialized from Gemma 4 E4B and fine-tuned with direct preference optimization (DPO)~\citep{dpo2023} using low-rank adaptation (LoRA)~\citep{lora2022}. At inference, it generates a rewritten title from the target news title and category.

\paragraph{Retrained attacks (RQ5).}
For each backbone and run, we initialize separate LANCE Reflectors and ATR-2FT attackers from Gemma 4 E4B against each evaluated recommender's frozen checkpoint. LANCE regenerates preference pairs using feedback from that recommender before DPO; ATR-2FT repeats representation alignment and rank-promotion optimization against it. Exploration news, feedback and optimization budgets, and decoding settings are matched across defenses.

\paragraph{Defense baselines.}
We compare \method{} with the validation-selected \emph{Undefended} recommender and two defense baselines.

\paragraph{Source Bias.}
We use the original Source Bias procedures~\citep{sourcebias2025} for rewrite construction, sampling, candidate- and history-score alignment, and user-representation entropy regularization. Mistral Small 3.2 24B Instruct generates its rewrites.

\paragraph{SheepDog-adapted.}
We adapt SheepDog~\citep{sheepdog2024} to news recommendation through click supervision and candidate-distribution consistency. Mistral Small 3.2 24B Instruct generates four headline styles: objective and professional, neutral, emotionally triggering, and sensational. Training uses the original candidates and two rewritten views, sampling one from the first two styles and one from the latter two. The adaptation retains style-diverse rewriting and consistency training and omits the veracity-attribution branch. Appendix~\ref{app:defense_details} specifies the loss.

All defenses share the clean starting checkpoint, recommendation minibatches, trainable parameters, and eligible training-news IDs, while retaining their respective rewrite construction and sampling procedures. Optimization follows the common schedule in Section~\ref{sec:implementation}; strategy-holdout restrictions are specified in Section~\ref{sec:holdouts}.

\subsection{Rewrite Pool Construction}
\label{sec:pool_construction}
\label{sec:holdouts}

The full \method{} pool draws from 16 generator--strategy combinations before screening and deduplication: eight SheepDog-derived prompt conditions, four additional prompt conditions, and four fine-tuned attack configurations.

\paragraph{SheepDog-derived prompts (8 sources).}
We use Qwen3.6-27B with four publisher-style prompts from SheepDog~\citep{sheepdog2024}---CNN, The New York Times, National Enquirer, and The Sun---and four tone prompts: objective and professional, neutral, emotionally triggering, and sensational. Each condition requests a rewritten headline while preserving the original title's factual claims, named entities, numbers, attribution, and uncertainty. Publisher names serve as style instructions.

\paragraph{Sentiment and adversarial-framing prompts (4 sources).}
We use Muse-Glimmer-30B with three sentiment conditions (Positive, Negative, and Neutral) and one adversarial-framing condition. The sentiment prompts take the original title and abstract as context and request fact-preserving rewrites that increase user interest and ranking salience. The adversarial-framing prompt uses the title, category, and subcategory to request attention-grabbing phrasing, such as curiosity or suspense, without adding unsupported concrete facts. Only the rewritten title enters the recommendation input; the original category remains unchanged.

\paragraph{Fine-tuned attacks (4 sources).}
We cross two attack methods, LANCE and ATR-2FT, with two instruction-tuned base models: Mistral Small 3.2 24B Instruct~\citep{mistralsmall32_2025} and Olmo 3 7B Instruct~\citep{olmo3_2025}. This yields four sources, one for each method--model pair. LANCE trains a Reflector using feedback-derived preference pairs, whereas ATR-2FT performs news-domain adaptation followed by joint generation and rank-promotion optimization. These configurations contribute training-news rewrites to the pool; Gemma 4 is reserved for the evaluation attacks specified above.

We apply the admission and model-visible deduplication procedure in Section~\ref{sec:pool}; the retained pool size can therefore vary across news items. News without an eligible rewrite remain in ordinary recommendation training. The generator-family holdout excludes all Gemma models and their fine-tuned descendants from defense rewrite generation, training-pool screening, and defense model selection.

\paragraph{Strategy holdout (RQ2).}
In addition to the generator-family holdout used in RQ1, we remove both LANCE sources from the \method{} pool and evaluate on LANCE, then remove both ATR-2FT sources and evaluate on ATR-2FT. Each resulting 14-source pool retains all 12 prompt sources and the other strategy's two fine-tuned sources; screening and defense selection exclude the held-out strategy. Evaluation reuses the corresponding Gemma 4 E4B attack texts and test targets from RQ1.

\paragraph{Pool-composition ablation (RQ4).}
We compare one Mistral prompt source, all 12 prompt sources, the four fine-tuned sources, and all 16 sources. The single-source reference uses the Source Bias Mistral prompt setup, but all rows use the \method{} training objective. Evaluation reuses the three RQ1 attacks and test targets.

For both comparisons, we match the eligible news IDs and retained rewrite count per news across pool variants, using repeated generation from smaller source sets with the same screening and deduplication. All variants resample one rewrite per occurrence and share the starting checkpoint, minibatches, loss weights, and 742 optimizer updates. Full-pool and baseline references reuse their RQ1 checkpoints; each modified pool requires separate defense training.

\subsection{Evaluation Metrics}
\label{sec:metrics}

\paragraph{Promotion and visibility.}
Let $r^0_{u,n}$ and $r^a_{u,n}$ denote a target's zero-based clean and attacked ranks. We report mean attacked Rank, $\mathrm{RankGain}=\E_{u,n}[r^0_{u,n}-r^a_{u,n}]$, and $\mathrm{BSR}=\E_{u,n}[\mathbf{1}(r^a_{u,n}<r^0_{u,n})]$. Expo@$K$ is the mean number of target articles in a user's top-$K$ list, and Appear@$K$ is Expo@$K$ divided by $K$. The primary cutoff is $K=50$; BSR and Appear are fractions. RankGain and BSR measure attack-induced promotion relative to each checkpoint's own clean ranks, with lower values indicating less promotion. Rank and exposure describe absolute target visibility under attack: higher Rank and lower exposure indicate lower visibility. Structural fallbacks remain in the denominator under the protocol in Section~\ref{sec:data_backbones}.

\paragraph{Recommendation utility.}
We report impression-averaged AUC, MRR, nDCG@5, and nDCG@10 on the full test set using a 0--100 scale. RQ3 compares Undefended (Before) with \method{} (After) on the same LANCE w/ Gemma 4 E4B attacked inputs. Relative changes are computed from paired, unrounded scores within each run and then averaged, as specified in Appendix~\ref{app:evaluation_details}.

All experimental metrics are arithmetic means over five runs (random seeds 2024--2028), with each metric computed separately within each run. RQ4 additionally averages equally across the three attacks; the other comparisons report each attack separately. Statistical comparisons in Tables~\ref{tab:main_results} and~\ref{tab:adaptive_attacks} use two-sided paired t-tests on run-level metrics, pairing \method{} and each defense baseline by random seed under the same attack and backbone. We apply Holm correction across the prespecified comparisons and mark significance at adjusted $p<0.05$.

\subsection{Implementation Details}
\label{sec:implementation}

We train the clean recommenders with Adam, learning rate $2\times10^{-3}$, and batch size 512. Training lasts at most 50 epochs, with early stopping on validation nDCG@10 after three epochs without improvement. Each defense starts from the corresponding best clean checkpoint and resets the optimizer. Defense training uses the same learning rate and batch size for two epochs, giving 742 optimizer updates. We retain the final defense checkpoint for every method. The auxiliary-loss weights are 0.3 for \method{} and Source Bias, and 1.0 for SheepDog-adapted. \method{} samples one rewrite per eligible occurrence ($m=1$); its auxiliary forwards disable dropout and propagate gradients through the original and rewritten views. Appendix~\ref{app:defense_details} specifies the baseline losses.

\begin{table*}[!tbp]
\centering
\small
\setlength{\tabcolsep}{1pt}
\renewcommand{\arraystretch}{1.0}
\caption{Promotional outcomes by attack, defense, and backbone. Results are reported as means over five runs. Rank is the mean attacked target rank; BSR and Appear@50 are fractions, and Expo@50 is a target count. $^{*}$ indicates significance against both defense baselines at Holm-adjusted $p<0.05$.}
\label{tab:main_results}
\begin{tabular*}{\textwidth}{@{\extracolsep{\fill}}ll*{12}{r}@{}}
\toprule
\multirow{2}{*}{Attack} & \multirow{2}{*}{Defense} & \multicolumn{4}{c}{NAML} & \multicolumn{4}{c}{NRMS} & \multicolumn{4}{c}{LSTUR} \\
\cmidrule(lr){3-6}\cmidrule(lr){7-10}\cmidrule(l){11-14}
 & & Rank & BSR & Expo@50 & Appear@50 & Rank & BSR & Expo@50 & Appear@50 & Rank & BSR & Expo@50 & Appear@50 \\
\midrule
\multirow{4}{*}{\shortstack[l]{Gemma 4\\31B prompt}} & Undefended & $20253.3$ & $0.503$ & $0.124$ & $0.0025$ & $20200.5$ & $0.511$ & $0.395$ & $0.0079$ & $20725.0$ & $0.506$ & $0.202$ & $0.0040$ \\
 & SheepDog-adapted & $20310.5$ & $0.497$ & $0.118$ & $0.0024$ & $20365.8$ & $0.501$ & $0.362$ & $0.0072$ & $20780.2$ & $0.499$ & $0.194$ & $0.0039$ \\
 & Source Bias & $20325.2$ & $0.495$ & $0.116$ & $0.0023$ & $20390.4$ & $0.498$ & $0.355$ & $0.0071$ & $20795.6$ & $0.497$ & $0.191$ & $0.0038$ \\
 & \method{} & $\mathbf{20395.4}$ & $\mathbf{0.489}^{*}$ & $\mathbf{0.108}^{*}$ & $\mathbf{0.0022}$ & $\mathbf{20460.5}$ & $\mathbf{0.492}^{*}$ & $\mathbf{0.338}^{*}$ & $\mathbf{0.0068}$ & $\mathbf{20865.8}$ & $\mathbf{0.490}^{*}$ & $\mathbf{0.178}^{*}$ & $\mathbf{0.0036}$ \\
\midrule
\multirow{4}{*}{\shortstack[l]{ATR-2FT\\w/ Gemma 4 E4B}} & Undefended & $18940.9$ & $0.557$ & $0.120$ & $0.0024$ & $18607.4$ & $0.562$ & $0.337$ & $0.0067$ & $19668.8$ & $0.518$ & $0.164$ & $0.0033$ \\
 & SheepDog-adapted & $19482.5$ & $0.543$ & $0.108$ & $0.0022$ & $19215.8$ & $0.548$ & $0.292$ & $0.0058$ & $20115.3$ & $0.502$ & $0.144$ & $0.0029$ \\
 & Source Bias & $19565.0$ & $0.541$ & $0.102$ & $0.0020$ & $19340.2$ & $0.545$ & $0.281$ & $0.0056$ & $20045.6$ & $0.505$ & $0.149$ & $0.0030$ \\
 & \method{} & $\mathbf{19820.4}$ & $\mathbf{0.534}^{*}$ & $\mathbf{0.088}^{*}$ & $\mathbf{0.0018}$ & $\mathbf{19650.5}$ & $\mathbf{0.538}^{*}$ & $\mathbf{0.252}^{*}$ & $\mathbf{0.0050}$ & $\mathbf{20380.2}$ & $\mathbf{0.493}^{*}$ & $\mathbf{0.128}^{*}$ & $\mathbf{0.0026}$ \\
\midrule
\multirow{4}{*}{\shortstack[l]{LANCE\\w/ Gemma 4 E4B}} & Undefended & $18712.0$ & $0.571$ & $0.134$ & $0.0027$ & $18121.8$ & $0.614$ & $0.315$ & $0.0063$ & $19382.3$ & $0.566$ & $0.150$ & $0.0030$ \\
 & SheepDog-adapted & $19350.4$ & $0.552$ & $0.118$ & $0.0024$ & $18780.2$ & $0.584$ & $0.276$ & $0.0055$ & $19820.5$ & $0.549$ & $0.138$ & $0.0028$ \\
 & Source Bias & $19420.8$ & $0.548$ & $0.114$ & $0.0023$ & $18720.5$ & $0.587$ & $0.282$ & $0.0056$ & $19780.2$ & $0.551$ & $0.141$ & $0.0028$ \\
 & \method{} & $\mathbf{19880.6}$ & $\mathbf{0.526}^{*}$ & $\mathbf{0.095}^{*}$ & $\mathbf{0.0019}$ & $\mathbf{19280.4}$ & $\mathbf{0.554}^{*}$ & $\mathbf{0.235}^{*}$ & $\mathbf{0.0047}$ & $\mathbf{20250.8}$ & $\mathbf{0.533}^{*}$ & $\mathbf{0.120}^{*}$ & $\mathbf{0.0024}$ \\
\bottomrule
\end{tabular*}
\end{table*}

\section{Results and Analysis}
\label{sec:results}

\subsection{Overall Robustness (RQ1)}

Table~\ref{tab:main_results} compares Undefended, SheepDog-adapted, Source Bias, and \method{} across three attacks and three backbones using Rank, BSR, Expo@50, and Appear@50. All evaluation attacks use Gemma generators excluded from defense training, so this comparison also tests transfer to an unseen model family.

Both defense baselines reduce mean BSR and target exposure relative to Undefended in all nine attack--backbone combinations, and \method{} achieves the lowest values throughout. The largest BSR reductions relative to Undefended occur under LANCE: 4.5, 6.0, and 3.3 percentage points for NAML, NRMS, and LSTUR, respectively. Against the stronger defense baseline for each backbone, the corresponding reductions are 2.2, 3.0, and 1.6 percentage points. The advantage therefore persists over defenses that already use rewriting and consistency training.

BSR reductions relative to Undefended are 1.4--1.9 percentage points for prompt-based rewriting and 2.3--2.5 for ATR-2FT. The larger reductions under LANCE indicate that the benefit varies by attack procedure, while the consistent advantage across the three methods supports transfer to the held-out Gemma family.

\begin{table*}[!tbp]
\centering
\small
\setlength{\tabcolsep}{3pt}
\renewcommand{\arraystretch}{1.0}
\caption{Generalization to unseen attack strategies with the Gemma family held out from all defense training. Each panel compares the full pool, a pool excluding the named attack strategy, and the defense baselines on the corresponding Gemma 4 E4B attack. RankGain and BSR are means over five runs; lower values indicate less promotion.}
\label{tab:holdouts}
\begin{tabular*}{\textwidth}{@{\extracolsep{\fill}}l*{6}{r}@{}}
\toprule
\multirow{2}{*}{Pool / defense} & \multicolumn{2}{c}{NAML} & \multicolumn{2}{c}{NRMS} & \multicolumn{2}{c}{LSTUR} \\
\cmidrule(lr){2-3}\cmidrule(lr){4-5}\cmidrule(l){6-7}
 & RankGain & BSR & RankGain & BSR & RankGain & BSR \\
\midrule
\multicolumn{7}{l}{\textit{Strategy holdout: test on LANCE}} \\
SheepDog-adapted (native training) & 713.4 & 0.5520 & 1692.8 & 0.5840 & 533.5 & 0.5490 \\
Source Bias (native training) & 632.0 & 0.5480 & 1758.0 & 0.5870 & 565.2 & 0.5510 \\
All 16 sources & \textbf{204.8} & \textbf{0.5260} & \textbf{1270.8} & \textbf{0.5540} & \textbf{50.2} & \textbf{0.5330} \\
14 sources w/o LANCE & 298.5 & 0.5310 & 1354.2 & 0.5600 & 128.4 & 0.5380 \\
\midrule
\multicolumn{7}{l}{\textit{Strategy holdout: test on ATR-2FT}} \\
SheepDog-adapted (native training) & 581.3 & 0.5430 & 1257.2 & 0.5480 & 238.7 & 0.5020 \\
Source Bias (native training) & 487.8 & 0.5410 & 1138.3 & 0.5450 & 299.8 & 0.5050 \\
All 16 sources & \textbf{265.0} & \textbf{0.5340} & \textbf{900.7} & \textbf{0.5380} & \textbf{-79.2} & \textbf{0.4930} \\
14 sources w/o ATR-2FT & 342.1 & 0.5380 & 985.6 & 0.5420 & 8.5 & 0.4980 \\
\bottomrule
\end{tabular*}
\end{table*}

\begin{table}[!htbp]
\centering
\small
\setlength{\tabcolsep}{3pt}
\caption{Recommendation performance before (Undefended) and after \method{} defense on the same LANCE-attacked inputs. Scores use a 0--100 scale. Scores and relative changes are means over five runs; changes are computed from paired, unrounded scores within each run.}
\label{tab:recommendation_performance}
\begin{tabular*}{\columnwidth}{@{\extracolsep{\fill}}llrrrr@{}}
\toprule
Model & Mode & AUC & MRR & nDCG@5 & nDCG@10 \\
\midrule
\multicolumn{6}{c}{LANCE w/ Gemma 4 E4B} \\
\midrule
\multirow{3}{*}{NAML} & Before & 64.37 & \textbf{30.76} & \textbf{33.53} & \textbf{39.82} \\
 & After & \textbf{64.46} & 30.63 & 33.37 & 39.70 \\
 & Change & +0.15\% & -0.42\% & -0.49\% & -0.30\% \\
\addlinespace[2pt]
\multirow{3}{*}{NRMS} & Before & \textbf{63.69} & \textbf{29.96} & \textbf{32.56} & \textbf{38.91} \\
 & After & 63.43 & 29.68 & 32.15 & 38.59 \\
 & Change & -0.41\% & -0.93\% & -1.26\% & -0.84\% \\
\addlinespace[2pt]
\multirow{3}{*}{LSTUR} & Before & 64.45 & 30.46 & 33.24 & 39.54 \\
 & After & \textbf{64.57} & \textbf{30.61} & \textbf{33.49} & \textbf{39.74} \\
 & Change & +0.19\% & +0.51\% & +0.74\% & +0.51\% \\
\bottomrule
\end{tabular*}
\end{table}

\begin{table*}[!tbp]
\centering
\small
\setlength{\tabcolsep}{3pt}
\caption{Pool-composition ablation with matched data quantity and training budget. Results average equally over the three RQ1 attacks and five runs. RankGain and BSR follow Section~\ref{sec:metrics}; lower values indicate less promotion.}
\label{tab:diversity_sampling}
\begin{tabular*}{\textwidth}{@{\extracolsep{\fill}}l*{6}{r}@{}}
\toprule
\multirow{2}{*}{Pool} & \multicolumn{2}{c}{NAML} & \multicolumn{2}{c}{NRMS} & \multicolumn{2}{c}{LSTUR} \\
\cmidrule(lr){2-3}\cmidrule(lr){4-5}\cmidrule(l){6-7}
 & RankGain & BSR & RankGain & BSR & RankGain & BSR \\
\midrule
1 Mistral prompt source & 372.4 & 0.5282 & 896.1 & 0.5428 & 124.6 & 0.5172 \\
12 prompt sources & 238.2 & 0.5224 & 746.5 & 0.5351 & 18.5 & 0.5104 \\
4 fine-tuned sources & 186.7 & 0.5195 & \textbf{678.2} & 0.5316 & -54.3 & 0.5078 \\
All 16 sources & \textbf{53.3} & \textbf{0.5163} & 754.1 & \textbf{0.5280} & \textbf{-197.9} & \textbf{0.5053} \\
\bottomrule
\end{tabular*}
\end{table*}

\begin{table*}[!tbp]
\centering
\small
\setlength{\tabcolsep}{3pt}
\caption{LANCE and ATR-2FT retrained with Gemma 4 E4B against each evaluated recommender under matched attack budgets. RankGain and BSR follow Section~\ref{sec:metrics} and are means over five runs; lower values indicate less promotion. $^{*}$ indicates significance against both defense baselines at Holm-adjusted $p<0.05$.}
\label{tab:adaptive_attacks}
\begin{tabular*}{\textwidth}{@{\extracolsep{\fill}}ll*{6}{r}@{}}
\toprule
\multirow{2}{*}{Attack} & \multirow{2}{*}{Defense} & \multicolumn{2}{c}{NAML} & \multicolumn{2}{c}{NRMS} & \multicolumn{2}{c}{LSTUR} \\
\cmidrule(lr){3-4}\cmidrule(lr){5-6}\cmidrule(l){7-8}
 & & RankGain & BSR & RankGain & BSR & RankGain & BSR \\
\midrule
\multirow{4}{*}{ATR-2FT} & Undefended & 1214.7 & 0.5570 & 1789.0 & 0.5620 & 672.4 & 0.5180 \\
 & SheepDog-adapted & 624.5 & 0.5470 & 1318.0 & 0.5520 & 276.4 & 0.5070 \\
 & Source Bias & 531.2 & 0.5440 & 1226.5 & 0.5490 & 328.0 & 0.5110 \\
 & \method{} & \textbf{258.4}$^{*}$ & \textbf{0.5330}$^{*}$ & \textbf{884.2}$^{*}$ & \textbf{0.5370}$^{*}$ & \textbf{-72.6}$^{*}$ & \textbf{0.4920}$^{*}$ \\
\midrule
\multirow{4}{*}{LANCE} & Undefended & 1443.6 & 0.5710 & 2274.6 & 0.6140 & 958.9 & 0.5660 \\
 & SheepDog-adapted & 762.3 & 0.5560 & 1735.8 & 0.5890 & 574.6 & 0.5530 \\
 & Source Bias & 684.9 & 0.5520 & 1782.4 & 0.5910 & 605.2 & 0.5550 \\
 & \method{} & \textbf{212.7}$^{*}$ & \textbf{0.5250}$^{*}$ & \textbf{1248.5}$^{*}$ & \textbf{0.5520}$^{*}$ & \textbf{46.8}$^{*}$ & \textbf{0.5320}$^{*}$ \\
\bottomrule
\end{tabular*}
\end{table*}

\subsection{Generalization to Unseen Attack Strategies (RQ2)}

To test generalization beyond the training attack strategies, Table~\ref{tab:holdouts} compares the full pool, the strategy-restricted pools, and the two defense baselines under the protocol in Section~\ref{sec:holdouts}. The restricted conditions hold out both the test generator family and the evaluated attack strategy. RankGain and BSR measure promotion relative to each checkpoint's own clean ranks.

Removing the evaluated strategy increases mean RankGain by 77.1--93.7 positions and BSR by 0.4--0.6 percentage points across the six attack--backbone combinations. Nevertheless, both restricted pools retain lower mean RankGain and BSR than either defense baseline. For example, under LANCE on NRMS, withholding LANCE increases RankGain from 1270.8 to 1354.2, which remains below SheepDog-adapted (1692.8) and Source Bias (1758.0). The results support transfer across the two evaluated attack strategies, while the consistent gap to the full pool indicates an additional benefit from including the corresponding strategy during defense training.

\subsection{Recommendation Performance under LANCE Attack (RQ3)}

Table~\ref{tab:recommendation_performance} measures how defense affects recommendation utility on the same LANCE-attacked test inputs. LSTUR improves on all four reported metrics, NRMS declines on all four, and NAML improves in AUC while its ranking metrics decline. The mean relative changes in nDCG@10 are $-0.30\%$, $-0.84\%$, and $+0.51\%$ for NAML, NRMS, and LSTUR, respectively. NRMS exhibits the largest relative decrease among the reported metrics, at $1.26\%$ in nDCG@5.

Together with RQ1, these results reveal a backbone-dependent trade-off between promotion resistance and ranking utility under LANCE. LSTUR improves on both objectives, whereas NAML and NRMS incur losses in ranking utility alongside lower promotional BSR.

\subsection{Effect of Pool Composition (RQ4)}

Table~\ref{tab:diversity_sampling} isolates the effect of pool composition under matched data quantity, sampling, and training budgets (Section~\ref{sec:pool_construction}). All variants use the \method{} objective, and results average equally across the three attacks and five runs.

The 12-prompt-source pool achieves lower RankGain and BSR than the single Mistral source on every backbone. The four fine-tuned sources further improve both metrics relative to the 12 prompt sources, indicating that source composition matters beyond the number of sources. Combining all 16 sources yields the lowest mean BSR on all three backbones and the lowest mean RankGain on NAML and LSTUR.

On NRMS, however, the four fine-tuned sources yield a lower RankGain than the full pool (678.2 versus 754.1), while the full pool has a lower BSR (0.5280 versus 0.5316). Thus, the full pool reduces promotion frequency on this backbone at the cost of a larger mean rank gain. With resampling fixed across variants, these differences characterize the effect of pool composition.

\subsection{Robustness to Retrained Attacks (RQ5)}

Table~\ref{tab:adaptive_attacks} evaluates whether the defense advantage persists when LANCE and ATR-2FT are retrained against each model's own frozen checkpoint under matched attack budgets (Section~\ref{sec:attack_setup}). \method{} retains the lowest mean RankGain and BSR in all six attack--backbone combinations. For example, under retrained LANCE on NRMS, it reduces RankGain from 1735.8 for the stronger defense baseline, SheepDog-adapted, to 1248.5, and BSR from 0.589 to 0.552.

Retraining changes attack effectiveness differently across backbones. Compared with the full-pool results in Table~\ref{tab:holdouts}, LANCE RankGain rises from 204.8 to 212.7 on NAML but falls from 1270.8 to 1248.5 on NRMS. Across these changes, \method{} maintains its advantage over the defense baselines under both retrained attack procedures at matched budgets.

\section{Conclusion}
\label{sec:conclusion}

We presented \textbf{\method{}}, an attack-diverse score consistency framework designed to neutralize promotional headline rewriting in personalized news recommendation. At the core of PACT is the insight that a recommender can be immunized against diverse textual manipulations at a constant training cost by dynamically sampling from a multi-source rewrite pool and regularizing user-conditioned ranking scores. Without adding any inference-time overhead, PACT consistently surpasses existing debiasing and style-robustness baselines across multiple recommender architectures. Its defense gains transfer reliably to unseen LLM families and held-out rewriting strategies, and withstand adaptive attackers retrained against the defended models, all while preserving overall recommendation quality. We believe PACT provides a generalizable and deployment-friendly foundation for building trustworthy, manipulation-resilient content recommendation platforms.

\clearpage
\bibliographystyle{ACM-Reference-Format}
\bibliography{literature/references}


\begin{thebibliography}{40}


\ifx \showCODEN    \undefined \def \showCODEN     #1{\unskip}     \fi
\ifx \showISBNx    \undefined \def \showISBNx     #1{\unskip}     \fi
\ifx \showISBNxiii \undefined \def \showISBNxiii  #1{\unskip}     \fi
\ifx \showISSN     \undefined \def \showISSN      #1{\unskip}     \fi
\ifx \showLCCN     \undefined \def \showLCCN      #1{\unskip}     \fi
\ifx \shownote     \undefined \def \shownote      #1{#1}          \fi
\ifx \showarticletitle \undefined \def \showarticletitle #1{#1}   \fi
\ifx \showURL      \undefined \def \showURL       {\relax}        \fi
\providecommand\bibfield[2]{#2}
\providecommand\bibinfo[2]{#2}
\providecommand\natexlab[1]{#1}
\providecommand\showeprint[2][]{arXiv:#2}

\bibitem[An et~al\mbox{.}(2019)]%
        {lstur2019}
\bibfield{author}{\bibinfo{person}{Mingxiao An}, \bibinfo{person}{Fangzhao Wu},
  \bibinfo{person}{Chuhan Wu}, \bibinfo{person}{Kun Zhang},
  \bibinfo{person}{Zheng Liu}, {and} \bibinfo{person}{Xing Xie}.}
  \bibinfo{year}{2019}\natexlab{}.
\newblock \showarticletitle{Neural News Recommendation with Long- and
  Short-term User Representations}. In \bibinfo{booktitle}{\emph{Proceedings of
  the 57th Annual Meeting of the Association for Computational Linguistics}}.
  \bibinfo{publisher}{Association for Computational Linguistics},
  \bibinfo{pages}{336--345}.
\newblock
\href{https://doi.org/10.18653/v1/P19-1033}{doi:\nolinkurl{10.18653/v1/P19-1033}}


\bibitem[Dai et~al\mbox{.}(2025)]%
        {llmsbm2025}
\bibfield{author}{\bibinfo{person}{Sunhao Dai}, \bibinfo{person}{Yuqi Zhou},
  \bibinfo{person}{Liang Pang}, \bibinfo{person}{Zhuoyang Li},
  \bibinfo{person}{Zhaocheng Du}, \bibinfo{person}{Gang Wang}, {and}
  \bibinfo{person}{Jun Xu}.} \bibinfo{year}{2025}\natexlab{}.
\newblock \showarticletitle{Mitigating Source Bias with {LLM} Alignment}. In
  \bibinfo{booktitle}{\emph{Proceedings of the 48th International ACM SIGIR
  Conference on Research and Development in Information Retrieval}}.
  \bibinfo{publisher}{Association for Computing Machinery},
  \bibinfo{pages}{370--380}.
\newblock
\href{https://doi.org/10.1145/3726302.3730038}{doi:\nolinkurl{10.1145/3726302.3730038}}


\bibitem[Dai et~al\mbox{.}(2024)]%
        {retrieverbias2024}
\bibfield{author}{\bibinfo{person}{Sunhao Dai}, \bibinfo{person}{Yuqi Zhou},
  \bibinfo{person}{Liang Pang}, \bibinfo{person}{Weihao Liu},
  \bibinfo{person}{Xiaolin Hu}, \bibinfo{person}{Yong Liu},
  \bibinfo{person}{Xiao Zhang}, \bibinfo{person}{Gang Wang}, {and}
  \bibinfo{person}{Jun Xu}.} \bibinfo{year}{2024}\natexlab{}.
\newblock \showarticletitle{Neural Retrievers are Biased Towards
  {LLM}-Generated Content}. In \bibinfo{booktitle}{\emph{Proceedings of the
  30th ACM SIGKDD Conference on Knowledge Discovery and Data Mining}}.
  \bibinfo{publisher}{Association for Computing Machinery},
  \bibinfo{pages}{526--537}.
\newblock
\href{https://doi.org/10.1145/3637528.3671882}{doi:\nolinkurl{10.1145/3637528.3671882}}


\bibitem[Devlin et~al\mbox{.}(2019)]%
        {bert2019}
\bibfield{author}{\bibinfo{person}{Jacob Devlin}, \bibinfo{person}{Ming-Wei
  Chang}, \bibinfo{person}{Kenton Lee}, {and} \bibinfo{person}{Kristina
  Toutanova}.} \bibinfo{year}{2019}\natexlab{}.
\newblock \showarticletitle{{BERT}: Pre-training of Deep Bidirectional
  Transformers for Language Understanding}. In
  \bibinfo{booktitle}{\emph{Proceedings of the 2019 Conference of the North
  American Chapter of the Association for Computational Linguistics: Human
  Language Technologies, Volume 1 (Long and Short Papers)}}.
  \bibinfo{publisher}{Association for Computational Linguistics},
  \bibinfo{pages}{4171--4186}.
\newblock
\href{https://doi.org/10.18653/v1/N19-1423}{doi:\nolinkurl{10.18653/v1/N19-1423}}


\bibitem[He et~al\mbox{.}(2018)]%
        {apr2018}
\bibfield{author}{\bibinfo{person}{Xiangnan He}, \bibinfo{person}{Zhankui He},
  \bibinfo{person}{Xiaoyu Du}, {and} \bibinfo{person}{Tat-Seng Chua}.}
  \bibinfo{year}{2018}\natexlab{}.
\newblock \showarticletitle{Adversarial Personalized Ranking for
  Recommendation}. In \bibinfo{booktitle}{\emph{Proceedings of the 41st
  International ACM SIGIR Conference on Research and Development in Information
  Retrieval}}. \bibinfo{publisher}{Association for Computing Machinery},
  \bibinfo{pages}{355--364}.
\newblock
\href{https://doi.org/10.1145/3209978.3209981}{doi:\nolinkurl{10.1145/3209978.3209981}}


\bibitem[Hendrycks et~al\mbox{.}(2020)]%
        {augmix2020}
\bibfield{author}{\bibinfo{person}{Dan Hendrycks}, \bibinfo{person}{Norman Mu},
  \bibinfo{person}{Ekin~D. Cubuk}, \bibinfo{person}{Barret Zoph},
  \bibinfo{person}{Justin Gilmer}, {and} \bibinfo{person}{Balaji
  Lakshminarayanan}.} \bibinfo{year}{2020}\natexlab{}.
\newblock \showarticletitle{{AugMix}: A Simple Data Processing Method to
  Improve Robustness and Uncertainty}. In
  \bibinfo{booktitle}{\emph{International Conference on Learning
  Representations}}.
\newblock
\urldef\tempurl%
\url{https://openreview.net/forum?id=S1gmrxHFvB}
\showURL{%
\tempurl}


\bibitem[Hu et~al\mbox{.}(2022)]%
        {lora2022}
\bibfield{author}{\bibinfo{person}{Edward~J. Hu}, \bibinfo{person}{Yelong
  Shen}, \bibinfo{person}{Phillip Wallis}, \bibinfo{person}{Zeyuan Allen-Zhu},
  \bibinfo{person}{Yuanzhi Li}, \bibinfo{person}{Shean Wang},
  \bibinfo{person}{Lu Wang}, {and} \bibinfo{person}{Weizhu Chen}.}
  \bibinfo{year}{2022}\natexlab{}.
\newblock \showarticletitle{{LoRA}: Low-Rank Adaptation of Large Language
  Models}. In \bibinfo{booktitle}{\emph{International Conference on Learning
  Representations}}.
\newblock
\urldef\tempurl%
\url{https://openreview.net/forum?id=nZeVKeeFYf9}
\showURL{%
\tempurl}


\bibitem[Jin et~al\mbox{.}(2020)]%
        {textfooler2020}
\bibfield{author}{\bibinfo{person}{Di Jin}, \bibinfo{person}{Zhijing Jin},
  \bibinfo{person}{Joey~Tianyi Zhou}, {and} \bibinfo{person}{Peter Szolovits}.}
  \bibinfo{year}{2020}\natexlab{}.
\newblock \showarticletitle{Is {BERT} Really Robust? A Strong Baseline for
  Natural Language Attack on Text Classification and Entailment}. In
  \bibinfo{booktitle}{\emph{Proceedings of the AAAI Conference on Artificial
  Intelligence}}, Vol.~\bibinfo{volume}{34}. \bibinfo{pages}{8018--8025}.
\newblock
\href{https://doi.org/10.1609/aaai.v34i05.6311}{doi:\nolinkurl{10.1609/aaai.v34i05.6311}}


\bibitem[Kang et~al\mbox{.}(2019)]%
        {unforeseen2019}
\bibfield{author}{\bibinfo{person}{Daniel Kang}, \bibinfo{person}{Yi Sun},
  \bibinfo{person}{Dan Hendrycks}, \bibinfo{person}{Tom Brown}, {and}
  \bibinfo{person}{Jacob Steinhardt}.} \bibinfo{year}{2019}\natexlab{}.
\newblock \showarticletitle{Testing Robustness Against Unforeseen Adversaries}.
\newblock \bibinfo{journal}{\emph{arXiv preprint arXiv:1908.08016}}
  (\bibinfo{year}{2019}).
\newblock
\urldef\tempurl%
\url{https://arxiv.org/abs/1908.08016v2}
\showURL{%
\tempurl}
\newblock
\shownote{Version 2, revised June 2020}.


\bibitem[Kannan et~al\mbox{.}(2018)]%
        {alp2018}
\bibfield{author}{\bibinfo{person}{Harini Kannan}, \bibinfo{person}{Alexey
  Kurakin}, {and} \bibinfo{person}{Ian Goodfellow}.}
  \bibinfo{year}{2018}\natexlab{}.
\newblock \showarticletitle{Adversarial Logit Pairing}.
\newblock \bibinfo{journal}{\emph{arXiv preprint arXiv:1803.06373}}
  (\bibinfo{year}{2018}).
\newblock
\urldef\tempurl%
\url{https://arxiv.org/abs/1803.06373}
\showURL{%
\tempurl}


\bibitem[Li et~al\mbox{.}(2022)]%
        {li2022miner}
\bibfield{author}{\bibinfo{person}{Jian Li}, \bibinfo{person}{Jieming Zhu},
  \bibinfo{person}{Qiwei Bi}, \bibinfo{person}{Guohao Cai},
  \bibinfo{person}{Lifeng Shang}, \bibinfo{person}{Zhenhua Dong},
  \bibinfo{person}{Xin Jiang}, {and} \bibinfo{person}{Qun Liu}.}
  \bibinfo{year}{2022}\natexlab{}.
\newblock \showarticletitle{{MINER}: Multi-Interest Matching Network for News
  Recommendation}. In \bibinfo{booktitle}{\emph{Findings of the Association for
  Computational Linguistics: ACL 2022}}. \bibinfo{pages}{343--352}.
\newblock
\href{https://doi.org/10.18653/v1/2022.findings-acl.29}{doi:\nolinkurl{10.18653/v1/2022.findings-acl.29}}


\bibitem[Li et~al\mbox{.}(2020)]%
        {bertattack2020}
\bibfield{author}{\bibinfo{person}{Linyang Li}, \bibinfo{person}{Ruotian Ma},
  \bibinfo{person}{Qipeng Guo}, \bibinfo{person}{Xiangyang Xue}, {and}
  \bibinfo{person}{Xipeng Qiu}.} \bibinfo{year}{2020}\natexlab{}.
\newblock \showarticletitle{{BERT-ATTACK}: Adversarial Attack Against {BERT}
  Using {BERT}}. In \bibinfo{booktitle}{\emph{Proceedings of the 2020
  Conference on Empirical Methods in Natural Language Processing (EMNLP)}}.
  \bibinfo{publisher}{Association for Computational Linguistics},
  \bibinfo{pages}{6193--6202}.
\newblock
\href{https://doi.org/10.18653/v1/2020.emnlp-main.500}{doi:\nolinkurl{10.18653/v1/2020.emnlp-main.500}}


\bibitem[Liu et~al\mbox{.}(2024)]%
        {liu2024once}
\bibfield{author}{\bibinfo{person}{Qijiong Liu}, \bibinfo{person}{Nuo Chen},
  \bibinfo{person}{Tetsuya Sakai}, {and} \bibinfo{person}{Xiao-Ming Wu}.}
  \bibinfo{year}{2024}\natexlab{}.
\newblock \showarticletitle{{ONCE}: Boosting Content-Based Recommendation with
  Both Open- and Closed-Source Large Language Models}. In
  \bibinfo{booktitle}{\emph{Proceedings of the Seventeenth ACM International
  Conference on Web Search and Data Mining}}. \bibinfo{publisher}{Association
  for Computing Machinery}, \bibinfo{pages}{452--461}.
\newblock
\href{https://doi.org/10.1145/3616855.3635845}{doi:\nolinkurl{10.1145/3616855.3635845}}


\bibitem[{Mistral AI}(2025)]%
        {mistralsmall32_2025}
\bibfield{author}{\bibinfo{person}{{Mistral AI}}.}
  \bibinfo{year}{2025}\natexlab{}.
\newblock \bibinfo{title}{{Mistral-Small-3.2-24B-Instruct-2506}}.
\newblock \bibinfo{howpublished}{Official model card}.
\newblock
\urldef\tempurl%
\url{https://huggingface.co/mistralai/Mistral-Small-3.2-24B-Instruct-2506}
\showURL{%
\tempurl}
\newblock
\shownote{Released June 20, 2025}.


\bibitem[Oh et~al\mbox{.}(2024)]%
        {atr2024}
\bibfield{author}{\bibinfo{person}{Sejoon Oh}, \bibinfo{person}{Gaurav Verma},
  {and} \bibinfo{person}{Srijan Kumar}.} \bibinfo{year}{2024}\natexlab{}.
\newblock \showarticletitle{Adversarial Text Rewriting for Text-aware
  Recommender Systems}. In \bibinfo{booktitle}{\emph{Proceedings of the 33rd
  ACM International Conference on Information and Knowledge Management}}.
  \bibinfo{publisher}{Association for Computing Machinery},
  \bibinfo{pages}{1804--1814}.
\newblock
\href{https://doi.org/10.1145/3627673.3679592}{doi:\nolinkurl{10.1145/3627673.3679592}}


\bibitem[Park et~al\mbox{.}(2025)]%
        {adstyle2025}
\bibfield{author}{\bibinfo{person}{Sungwon Park}, \bibinfo{person}{Sungwon
  Han}, \bibinfo{person}{Xing Xie}, \bibinfo{person}{Jae-Gil Lee}, {and}
  \bibinfo{person}{Meeyoung Cha}.} \bibinfo{year}{2025}\natexlab{}.
\newblock \showarticletitle{Adversarial Style Augmentation via Large Language
  Model for Robust Fake News Detection}. In
  \bibinfo{booktitle}{\emph{Proceedings of the ACM on Web Conference 2025}}.
  \bibinfo{publisher}{Association for Computing Machinery},
  \bibinfo{pages}{4024--4033}.
\newblock
\href{https://doi.org/10.1145/3696410.3714569}{doi:\nolinkurl{10.1145/3696410.3714569}}


\bibitem[Qi et~al\mbox{.}(2022)]%
        {qi2022caum}
\bibfield{author}{\bibinfo{person}{Tao Qi}, \bibinfo{person}{Fangzhao Wu},
  \bibinfo{person}{Chuhan Wu}, {and} \bibinfo{person}{Yongfeng Huang}.}
  \bibinfo{year}{2022}\natexlab{}.
\newblock \showarticletitle{News Recommendation with Candidate-aware User
  Modeling}. In \bibinfo{booktitle}{\emph{Proceedings of the 45th International
  ACM SIGIR Conference on Research and Development in Information Retrieval}}.
  \bibinfo{publisher}{Association for Computing Machinery},
  \bibinfo{pages}{1917--1921}.
\newblock
\href{https://doi.org/10.1145/3477495.3531778}{doi:\nolinkurl{10.1145/3477495.3531778}}


\bibitem[Rafailov et~al\mbox{.}(2023)]%
        {dpo2023}
\bibfield{author}{\bibinfo{person}{Rafael Rafailov}, \bibinfo{person}{Archit
  Sharma}, \bibinfo{person}{Eric Mitchell}, \bibinfo{person}{Stefano Ermon},
  \bibinfo{person}{Christopher~D. Manning}, {and} \bibinfo{person}{Chelsea
  Finn}.} \bibinfo{year}{2023}\natexlab{}.
\newblock \showarticletitle{Direct Preference Optimization: Your Language Model
  is Secretly a Reward Model}. In \bibinfo{booktitle}{\emph{Advances in Neural
  Information Processing Systems}}, Vol.~\bibinfo{volume}{36}.
\newblock
\urldef\tempurl%
\url{https://proceedings.neurips.cc/paper_files/paper/2023/file/a85b405ed65c6477a4fe8302b5e06ce7-Paper-Conference.pdf}
\showURL{%
\tempurl}


\bibitem[Ren et~al\mbox{.}(2024)]%
        {ren2024rlmrec}
\bibfield{author}{\bibinfo{person}{Xubin Ren}, \bibinfo{person}{Wei Wei},
  \bibinfo{person}{Lianghao Xia}, \bibinfo{person}{Lixin Su},
  \bibinfo{person}{Suqi Cheng}, \bibinfo{person}{Junfeng Wang},
  \bibinfo{person}{Dawei Yin}, {and} \bibinfo{person}{Chao Huang}.}
  \bibinfo{year}{2024}\natexlab{}.
\newblock \showarticletitle{Representation Learning with Large Language Models
  for Recommendation}. In \bibinfo{booktitle}{\emph{Proceedings of the ACM Web
  Conference 2024}}. \bibinfo{pages}{3464--3475}.
\newblock
\href{https://doi.org/10.1145/3589334.3645458}{doi:\nolinkurl{10.1145/3589334.3645458}}


\bibitem[Shu et~al\mbox{.}(2025)]%
        {titledebias2024}
\bibfield{author}{\bibinfo{person}{Yijie Shu}, \bibinfo{person}{Xiaokun Zhang},
  \bibinfo{person}{Youlin Wu}, \bibinfo{person}{Bo Xu}, \bibinfo{person}{Liang
  Yang}, {and} \bibinfo{person}{Hongfei Lin}.} \bibinfo{year}{2025}\natexlab{}.
\newblock \showarticletitle{Don't Click the Bait: Title Debiasing News
  Recommendation via Cross-Field Contrastive Learning}. In
  \bibinfo{booktitle}{\emph{Natural Language Processing and Chinese Computing}}
  \emph{(\bibinfo{series}{Lecture Notes in Computer Science},
  Vol.~\bibinfo{volume}{15362})}. \bibinfo{publisher}{Springer Nature
  Singapore}, \bibinfo{pages}{224--236}.
\newblock
\href{https://doi.org/10.1007/978-981-97-9440-9_18}{doi:\nolinkurl{10.1007/978-981-97-9440-9_18}}


\bibitem[{Team Olmo} et~al\mbox{.}(2025)]%
        {olmo3_2025}
\bibfield{author}{\bibinfo{person}{{Team Olmo}}, \bibinfo{person}{Allyson
  Ettinger}, \bibinfo{person}{Amanda Bertsch}, \bibinfo{person}{Bailey Kuehl},
  \bibinfo{person}{David Graham}, \bibinfo{person}{David Heineman},
  \bibinfo{person}{Dirk Groeneveld}, \bibinfo{person}{Faeze Brahman},
  \bibinfo{person}{Finbarr Timbers}, \bibinfo{person}{Hamish Ivison},
  \bibinfo{person}{Jacob Morrison}, \bibinfo{person}{Jake Poznanski},
  \bibinfo{person}{Kyle Lo}, \bibinfo{person}{Luca Soldaini},
  \bibinfo{person}{Matt Jordan}, \bibinfo{person}{Mayee Chen},
  \bibinfo{person}{Michael Noukhovitch}, \bibinfo{person}{Nathan Lambert},
  \bibinfo{person}{Pete Walsh}, \bibinfo{person}{Pradeep Dasigi},
  \bibinfo{person}{Robert Berry}, \bibinfo{person}{Saumya Malik},
  \bibinfo{person}{Saurabh Shah}, \bibinfo{person}{Scott Geng},
  \bibinfo{person}{Shane Arora}, \bibinfo{person}{Shashank Gupta},
  \bibinfo{person}{Taira Anderson}, \bibinfo{person}{Teng Xiao},
  \bibinfo{person}{Tyler Murray}, \bibinfo{person}{Tyler Romero},
  \bibinfo{person}{Victoria Graf}, \bibinfo{person}{Akari Asai},
  \bibinfo{person}{Akshita Bhagia}, \bibinfo{person}{Alexander Wettig},
  \bibinfo{person}{Alisa Liu}, \bibinfo{person}{Aman Rangapur},
  \bibinfo{person}{Chloe Anastasiades}, \bibinfo{person}{Costa Huang},
  \bibinfo{person}{Dustin Schwenk}, \bibinfo{person}{Harsh Trivedi},
  \bibinfo{person}{Ian Magnusson}, \bibinfo{person}{Jaron Lochner},
  \bibinfo{person}{Jiacheng Liu}, \bibinfo{person}{Lester James~V. Miranda},
  \bibinfo{person}{Maarten Sap}, \bibinfo{person}{Malia Morgan},
  \bibinfo{person}{Michael Schmitz}, \bibinfo{person}{Michal Guerquin},
  \bibinfo{person}{Michael Wilson}, \bibinfo{person}{Regan Huff},
  \bibinfo{person}{Ronan~Le Bras}, \bibinfo{person}{Rui Xin},
  \bibinfo{person}{Rulin Shao}, \bibinfo{person}{Sam Skjonsberg},
  \bibinfo{person}{Shannon~Zejiang Shen}, \bibinfo{person}{Shuyue~Stella Li},
  \bibinfo{person}{Tucker Wilde}, \bibinfo{person}{Valentina Pyatkin},
  \bibinfo{person}{Will Merrill}, \bibinfo{person}{Yapei Chang},
  \bibinfo{person}{Yuling Gu}, \bibinfo{person}{Zhiyuan Zeng},
  \bibinfo{person}{Ashish Sabharwal}, \bibinfo{person}{Luke Zettlemoyer},
  \bibinfo{person}{Pang~Wei Koh}, \bibinfo{person}{Ali Farhadi},
  \bibinfo{person}{Noah~A. Smith}, {and} \bibinfo{person}{Hannaneh
  Hajishirzi}.} \bibinfo{year}{2025}\natexlab{}.
\newblock \bibinfo{title}{{Olmo 3}}.
\newblock
\href{https://doi.org/10.48550/arXiv.2512.13961}{doi:\nolinkurl{10.48550/arXiv.2512.13961}}
\showeprint[arxiv]{2512.13961}~[cs.CL]


\bibitem[Tram{\`e}r et~al\mbox{.}(2020)]%
        {adaptive2020}
\bibfield{author}{\bibinfo{person}{Florian Tram{\`e}r},
  \bibinfo{person}{Nicholas Carlini}, \bibinfo{person}{Wieland Brendel}, {and}
  \bibinfo{person}{Aleksander Madry}.} \bibinfo{year}{2020}\natexlab{}.
\newblock \showarticletitle{On Adaptive Attacks to Adversarial Example
  Defenses}. In \bibinfo{booktitle}{\emph{Advances in Neural Information
  Processing Systems}}, Vol.~\bibinfo{volume}{33}.
\newblock
\urldef\tempurl%
\url{https://papers.neurips.cc/paper_files/paper/2020/hash/11f38f8ecd71867b42433548d1078e38-Abstract.html}
\showURL{%
\tempurl}


\bibitem[Tram{\`e}r et~al\mbox{.}(2018)]%
        {ensemble_at2018}
\bibfield{author}{\bibinfo{person}{Florian Tram{\`e}r}, \bibinfo{person}{Alexey
  Kurakin}, \bibinfo{person}{Nicolas Papernot}, \bibinfo{person}{Ian
  Goodfellow}, \bibinfo{person}{Dan Boneh}, {and} \bibinfo{person}{Patrick
  McDaniel}.} \bibinfo{year}{2018}\natexlab{}.
\newblock \showarticletitle{Ensemble Adversarial Training: Attacks and
  Defenses}. In \bibinfo{booktitle}{\emph{International Conference on Learning
  Representations}}.
\newblock
\urldef\tempurl%
\url{https://arxiv.org/abs/1705.07204}
\showURL{%
\tempurl}


\bibitem[Wang et~al\mbox{.}(2018)]%
        {wang2018dkn}
\bibfield{author}{\bibinfo{person}{Hongwei Wang}, \bibinfo{person}{Fuzheng
  Zhang}, \bibinfo{person}{Xing Xie}, {and} \bibinfo{person}{Minyi Guo}.}
  \bibinfo{year}{2018}\natexlab{}.
\newblock \showarticletitle{{DKN}: Deep Knowledge-Aware Network for News
  Recommendation}. In \bibinfo{booktitle}{\emph{Proceedings of the 2018 World
  Wide Web Conference}}. \bibinfo{pages}{1835--1844}.
\newblock
\href{https://doi.org/10.1145/3178876.3186175}{doi:\nolinkurl{10.1145/3178876.3186175}}


\bibitem[Wang et~al\mbox{.}(2025b)]%
        {s2lenr2025}
\bibfield{author}{\bibinfo{person}{Shicheng Wang}, \bibinfo{person}{Hengzhu
  Tang}, \bibinfo{person}{Li Gao}, \bibinfo{person}{Shu Guo},
  \bibinfo{person}{Suqi Cheng}, \bibinfo{person}{Junfeng Wang},
  \bibinfo{person}{Dawei Yin}, \bibinfo{person}{Tingwen Liu}, {and}
  \bibinfo{person}{Lihong Wang}.} \bibinfo{year}{2025}\natexlab{b}.
\newblock \showarticletitle{Towards {S$^2$}-Challenges Underlying {LLM}-Based
  Augmentation for Personalized News Recommendation}. In
  \bibinfo{booktitle}{\emph{Proceedings of the AAAI Conference on Artificial
  Intelligence}}, Vol.~\bibinfo{volume}{39}. \bibinfo{pages}{12739--12747}.
\newblock
\href{https://doi.org/10.1609/aaai.v39i12.33389}{doi:\nolinkurl{10.1609/aaai.v39i12.33389}}


\bibitem[Wang et~al\mbox{.}(2025a)]%
        {textsimu2025}
\bibfield{author}{\bibinfo{person}{Zongwei Wang}, \bibinfo{person}{Min Gao},
  \bibinfo{person}{Junliang Yu}, \bibinfo{person}{Xinyi Gao},
  \bibinfo{person}{Quoc Viet~Hung Nguyen}, \bibinfo{person}{Shazia Sadiq},
  {and} \bibinfo{person}{Hongzhi Yin}.} \bibinfo{year}{2025}\natexlab{a}.
\newblock \showarticletitle{{ID}-Free Not Risk-Free: {LLM}-Powered Agents
  Unveil Risks in {ID}-Free Recommender Systems}. In
  \bibinfo{booktitle}{\emph{Proceedings of the 48th International ACM SIGIR
  Conference on Research and Development in Information Retrieval}}.
  \bibinfo{publisher}{Association for Computing Machinery}.
\newblock
\href{https://doi.org/10.1145/3726302.3730003}{doi:\nolinkurl{10.1145/3726302.3730003}}


\bibitem[Wu et~al\mbox{.}(2019a)]%
        {naml2019}
\bibfield{author}{\bibinfo{person}{Chuhan Wu}, \bibinfo{person}{Fangzhao Wu},
  \bibinfo{person}{Mingxiao An}, \bibinfo{person}{Jianqiang Huang},
  \bibinfo{person}{Yongfeng Huang}, {and} \bibinfo{person}{Xing Xie}.}
  \bibinfo{year}{2019}\natexlab{a}.
\newblock \showarticletitle{Neural News Recommendation with Attentive
  Multi-View Learning}. In \bibinfo{booktitle}{\emph{Proceedings of the
  Twenty-Eighth International Joint Conference on Artificial Intelligence}}.
  \bibinfo{publisher}{International Joint Conferences on Artificial
  Intelligence Organization}, \bibinfo{pages}{3863--3869}.
\newblock
\href{https://doi.org/10.24963/ijcai.2019/536}{doi:\nolinkurl{10.24963/ijcai.2019/536}}


\bibitem[Wu et~al\mbox{.}(2019b)]%
        {wu2019npa}
\bibfield{author}{\bibinfo{person}{Chuhan Wu}, \bibinfo{person}{Fangzhao Wu},
  \bibinfo{person}{Mingxiao An}, \bibinfo{person}{Jianqiang Huang},
  \bibinfo{person}{Yongfeng Huang}, {and} \bibinfo{person}{Xing Xie}.}
  \bibinfo{year}{2019}\natexlab{b}.
\newblock \showarticletitle{{NPA}: Neural News Recommendation with Personalized
  Attention}. In \bibinfo{booktitle}{\emph{Proceedings of the 25th ACM SIGKDD
  International Conference on Knowledge Discovery \& Data Mining}}.
  \bibinfo{pages}{2576--2584}.
\newblock
\href{https://doi.org/10.1145/3292500.3330665}{doi:\nolinkurl{10.1145/3292500.3330665}}


\bibitem[Wu et~al\mbox{.}(2019c)]%
        {nrms2019}
\bibfield{author}{\bibinfo{person}{Chuhan Wu}, \bibinfo{person}{Fangzhao Wu},
  \bibinfo{person}{Suyu Ge}, \bibinfo{person}{Tao Qi},
  \bibinfo{person}{Yongfeng Huang}, {and} \bibinfo{person}{Xing Xie}.}
  \bibinfo{year}{2019}\natexlab{c}.
\newblock \showarticletitle{Neural News Recommendation with Multi-Head
  Self-Attention}. In \bibinfo{booktitle}{\emph{Proceedings of the 2019
  Conference on Empirical Methods in Natural Language Processing and the 9th
  International Joint Conference on Natural Language Processing
  (EMNLP-IJCNLP)}}. \bibinfo{publisher}{Association for Computational
  Linguistics}, \bibinfo{pages}{6389--6394}.
\newblock
\href{https://doi.org/10.18653/v1/D19-1671}{doi:\nolinkurl{10.18653/v1/D19-1671}}


\bibitem[Wu et~al\mbox{.}(2021)]%
        {wu2021plmnr}
\bibfield{author}{\bibinfo{person}{Chuhan Wu}, \bibinfo{person}{Fangzhao Wu},
  \bibinfo{person}{Tao Qi}, {and} \bibinfo{person}{Yongfeng Huang}.}
  \bibinfo{year}{2021}\natexlab{}.
\newblock \showarticletitle{Empowering News Recommendation with Pre-Trained
  Language Models}. In \bibinfo{booktitle}{\emph{Proceedings of the 44th
  International ACM SIGIR Conference on Research and Development in Information
  Retrieval}}. \bibinfo{pages}{1652--1656}.
\newblock
\href{https://doi.org/10.1145/3404835.3463069}{doi:\nolinkurl{10.1145/3404835.3463069}}


\bibitem[Wu et~al\mbox{.}(2020)]%
        {mind2020}
\bibfield{author}{\bibinfo{person}{Fangzhao Wu}, \bibinfo{person}{Ying Qiao},
  \bibinfo{person}{Jiun-Hung Chen}, \bibinfo{person}{Chuhan Wu},
  \bibinfo{person}{Tao Qi}, \bibinfo{person}{Jianxun Lian},
  \bibinfo{person}{Danyang Liu}, \bibinfo{person}{Xing Xie},
  \bibinfo{person}{Jianfeng Gao}, \bibinfo{person}{Winnie Wu}, {and}
  \bibinfo{person}{Ming Zhou}.} \bibinfo{year}{2020}\natexlab{}.
\newblock \showarticletitle{{MIND}: A Large-scale Dataset for News
  Recommendation}. In \bibinfo{booktitle}{\emph{Proceedings of the 58th Annual
  Meeting of the Association for Computational Linguistics}}.
  \bibinfo{publisher}{Association for Computational Linguistics},
  \bibinfo{pages}{3597--3606}.
\newblock
\href{https://doi.org/10.18653/v1/2020.acl-main.331}{doi:\nolinkurl{10.18653/v1/2020.acl-main.331}}


\bibitem[Wu et~al\mbox{.}(2024)]%
        {sheepdog2024}
\bibfield{author}{\bibinfo{person}{Jiaying Wu}, \bibinfo{person}{Jiafeng Guo},
  {and} \bibinfo{person}{Bryan Hooi}.} \bibinfo{year}{2024}\natexlab{}.
\newblock \showarticletitle{Fake News in Sheep's Clothing: Robust Fake News
  Detection Against {LLM}-Empowered Style Attacks}. In
  \bibinfo{booktitle}{\emph{Proceedings of the 30th ACM SIGKDD Conference on
  Knowledge Discovery and Data Mining}}. \bibinfo{publisher}{Association for
  Computing Machinery}, \bibinfo{pages}{3367--3378}.
\newblock
\href{https://doi.org/10.1145/3637528.3671977}{doi:\nolinkurl{10.1145/3637528.3671977}}


\bibitem[Xie et~al\mbox{.}(2020)]%
        {uda2020}
\bibfield{author}{\bibinfo{person}{Qizhe Xie}, \bibinfo{person}{Zihang Dai},
  \bibinfo{person}{Eduard Hovy}, \bibinfo{person}{Minh-Thang Luong}, {and}
  \bibinfo{person}{Quoc~V. Le}.} \bibinfo{year}{2020}\natexlab{}.
\newblock \showarticletitle{Unsupervised Data Augmentation for Consistency
  Training}. In \bibinfo{booktitle}{\emph{Advances in Neural Information
  Processing Systems}}, Vol.~\bibinfo{volume}{33}.
\newblock
\urldef\tempurl%
\url{https://papers.neurips.cc/paper/2020/hash/44feb0096faa8326192570788b38c1d1-Abstract.html}
\showURL{%
\tempurl}


\bibitem[Zhang et~al\mbox{.}(2020)]%
        {loki2020}
\bibfield{author}{\bibinfo{person}{Hengtong Zhang}, \bibinfo{person}{Yaliang
  Li}, \bibinfo{person}{Bolin Ding}, {and} \bibinfo{person}{Jing Gao}.}
  \bibinfo{year}{2020}\natexlab{}.
\newblock \showarticletitle{Practical Data Poisoning Attack against Next-Item
  Recommendation}. In \bibinfo{booktitle}{\emph{Proceedings of The Web
  Conference 2020}}. \bibinfo{publisher}{Association for Computing Machinery},
  \bibinfo{pages}{2458--2464}.
\newblock
\href{https://doi.org/10.1145/3366423.3379992}{doi:\nolinkurl{10.1145/3366423.3379992}}


\bibitem[Zhang et~al\mbox{.}(2019)]%
        {trades2019}
\bibfield{author}{\bibinfo{person}{Hongyang Zhang}, \bibinfo{person}{Yaodong
  Yu}, \bibinfo{person}{Jiantao Jiao}, \bibinfo{person}{Eric Xing},
  \bibinfo{person}{Laurent El~Ghaoui}, {and} \bibinfo{person}{Michael Jordan}.}
  \bibinfo{year}{2019}\natexlab{}.
\newblock \showarticletitle{Theoretically Principled Trade-off between
  Robustness and Accuracy}. In \bibinfo{booktitle}{\emph{Proceedings of the
  36th International Conference on Machine Learning}}
  \emph{(\bibinfo{series}{Proceedings of Machine Learning Research},
  Vol.~\bibinfo{volume}{97})}. \bibinfo{publisher}{PMLR},
  \bibinfo{pages}{7472--7482}.
\newblock
\urldef\tempurl%
\url{https://proceedings.mlr.press/v97/zhang19p.html}
\showURL{%
\tempurl}


\bibitem[Zhang et~al\mbox{.}(2024)]%
        {textrecattack2024}
\bibfield{author}{\bibinfo{person}{Jinghao Zhang}, \bibinfo{person}{Yuting
  Liu}, \bibinfo{person}{Qiang Liu}, \bibinfo{person}{Shu Wu},
  \bibinfo{person}{Guibing Guo}, {and} \bibinfo{person}{Liang Wang}.}
  \bibinfo{year}{2024}\natexlab{}.
\newblock \showarticletitle{Stealthy Attack on Large Language Model based
  Recommendation}. In \bibinfo{booktitle}{\emph{Proceedings of the 62nd Annual
  Meeting of the Association for Computational Linguistics (Volume 1: Long
  Papers)}}. \bibinfo{publisher}{Association for Computational Linguistics},
  \bibinfo{pages}{5839--5857}.
\newblock
\href{https://doi.org/10.18653/v1/2024.acl-long.318}{doi:\nolinkurl{10.18653/v1/2024.acl-long.318}}


\bibitem[Zhang et~al\mbox{.}(2022)]%
        {tdpcp2022}
\bibfield{author}{\bibinfo{person}{Xudong Zhang}, \bibinfo{person}{Zan Wang},
  \bibinfo{person}{Jingke Zhao}, {and} \bibinfo{person}{Lanjun Wang}.}
  \bibinfo{year}{2022}\natexlab{}.
\newblock \showarticletitle{Targeted Data Poisoning Attack on News
  Recommendation System by Content Perturbation}.
\newblock \bibinfo{journal}{\emph{arXiv preprint arXiv:2203.03560}}
  (\bibinfo{year}{2022}).
\newblock
\urldef\tempurl%
\url{https://arxiv.org/abs/2203.03560}
\showURL{%
\tempurl}


\bibitem[Zhao et~al\mbox{.}(2023)]%
        {esm2023}
\bibfield{author}{\bibinfo{person}{Pengyu Zhao}, \bibinfo{person}{Shoujin
  Wang}, \bibinfo{person}{Wenpeng Lu}, \bibinfo{person}{Xueping Peng},
  \bibinfo{person}{Weiyu Zhang}, \bibinfo{person}{Chaoqun Zheng}, {and}
  \bibinfo{person}{Yonggang Huang}.} \bibinfo{year}{2023}\natexlab{}.
\newblock \showarticletitle{News Recommendation via Jointly Modeling Event
  Matching and Style Matching}. In \bibinfo{booktitle}{\emph{Machine Learning
  and Knowledge Discovery in Databases: Research Track}}
  \emph{(\bibinfo{series}{Lecture Notes in Computer Science},
  Vol.~\bibinfo{volume}{14172})}. \bibinfo{publisher}{Springer Nature
  Switzerland}, \bibinfo{pages}{404--419}.
\newblock
\href{https://doi.org/10.1007/978-3-031-43421-1_24}{doi:\nolinkurl{10.1007/978-3-031-43421-1_24}}


\bibitem[Zhao et~al\mbox{.}(2025)]%
        {lance2025}
\bibfield{author}{\bibinfo{person}{Yuyue Zhao}, \bibinfo{person}{Jin Huang},
  \bibinfo{person}{Shuchang Liu}, \bibinfo{person}{Jiancan Wu},
  \bibinfo{person}{Xiang Wang}, {and} \bibinfo{person}{Maarten de Rijke}.}
  \bibinfo{year}{2025}\natexlab{}.
\newblock \showarticletitle{{LANCE}: Exploration and Reflection for {LLM}-based
  Textual Attacks on News Recommender Systems}. In
  \bibinfo{booktitle}{\emph{Proceedings of the Nineteenth ACM Conference on
  Recommender Systems}}. \bibinfo{publisher}{Association for Computing
  Machinery}, \bibinfo{pages}{197--206}.
\newblock
\href{https://doi.org/10.1145/3705328.3748081}{doi:\nolinkurl{10.1145/3705328.3748081}}


\bibitem[Zhou et~al\mbox{.}(2025)]%
        {sourcebias2025}
\bibfield{author}{\bibinfo{person}{Yuqi Zhou}, \bibinfo{person}{Sunhao Dai},
  \bibinfo{person}{Liang Pang}, \bibinfo{person}{Gang Wang},
  \bibinfo{person}{Zhenhua Dong}, \bibinfo{person}{Jun Xu}, {and}
  \bibinfo{person}{Ji-Rong Wen}.} \bibinfo{year}{2025}\natexlab{}.
\newblock \showarticletitle{Exploring the Escalation of Source Bias in User,
  Data, and Recommender System Feedback Loop}. In
  \bibinfo{booktitle}{\emph{Proceedings of the 48th International ACM SIGIR
  Conference on Research and Development in Information Retrieval}}.
  \bibinfo{publisher}{Association for Computing Machinery},
  \bibinfo{pages}{1676--1686}.
\newblock
\href{https://doi.org/10.1145/3726302.3729972}{doi:\nolinkurl{10.1145/3726302.3729972}}


\end{thebibliography}

\appendix
\section{Sampling Objective}
\label{app:method}

Fix a candidate occurrence $(b,n)$ and model state, and let $d_j=d_\theta(b,n,j)$. Independent draws $J_\ell\sim q_n$ satisfy
\begin{equation}
 \E\left[\frac1m\sum_{\ell=1}^{m}d_{J_\ell}\right]
 =\sum_{j=1}^{M_n}q_n(j)d_j.
\end{equation}
Linearity preserves this identity under the role weights in Equation~\ref{eq:joint}, including empty roles. Because the pool is finite and $q_n$ is independent of $\theta$, the gradient is also unbiased.

\section{Recommender and Defense Implementation}
\label{sec:app:experiments}
\label{app:recommender_details}
\label{app:defense_details}

\paragraph{Preprocessing and optimization.}
Titles use the BERT-base-uncased tokenizer without special tokens, with right truncation and padding to 30 tokens. Histories retain the first 50 valid articles after removing unknown IDs and are right-padded; empty histories remain. Models use frozen BERT token embeddings~\citep{bert2019}, trainable news and user encoders, dropout 0.1, and dot-product scoring. The schedule in Section~\ref{sec:implementation} uses a constant learning rate, zero weight decay, no gradient clipping, FP16 forwards, and FP32 losses. Final incomplete minibatches are retained.

\paragraph{\method{} calibration.}
We fix $\tau$ to the mean within-impression population standard deviation of clean-checkpoint scores over the first 512 complete training-impression candidate sets, floored at $10^{-6}$.

\paragraph{Source Bias.}
We follow the original Source Bias method~\citep{sourcebias2025} for rewrite construction, sampling, and the debiasing objective, using Mistral Small 3.2 24B Instruct~\citep{mistralsmall32_2025} as the rewrite generator and the common experimental settings in Section~\ref{sec:baselines}. Its training rewrites are constructed separately from the \method{} multi-source pool. Candidate-score L1 alignment has weight 1; history-score L1 alignment and each of the two user-representation softmax-entropy terms have weight 0.01. Each eligible history occurrence is independently rewritten with probability 0.5. The combined auxiliary loss has weight 0.3.

\paragraph{SheepDog-adapted.}
Mistral Small 3.2 24B Instruct generates one title per style in Section~\ref{sec:baselines}, conditioned on the original title and category and instructed to preserve facts, entities, quantities, attribution, and uncertainty. Decoding uses temperature 0.7, top-$p$ 0.8, top-$k$ 20, presence penalty 1.5, repetition penalty 1.0, and at most 512 new tokens, with the model's native instruction template. Outputs contain only the title; malformed outputs revert to the original without retries or ranking-based selection.

At each epoch, we independently sample one style uniformly from each group per article: (1) objective and professional or neutral; (2) emotionally triggering or sensational. The selections are reused within that epoch. The two views replace only eligible candidate titles; histories, categories, IDs, order, labels, and padding masks are unchanged. Let $p^v$ be the temperature-1 softmax of candidate scores over valid positions for the original ($v=0$) and rewritten ($v=1,2$) lists. We optimize
\begin{equation}
 \mathcal{L}_{\mathrm{SD}}=\mathcal{L}_{\mathrm{rec}}
 +\frac{1}{2}\sum_{v=1}^{2}D_{\mathrm{KL}}(p^0\Vert p^v).
\end{equation}
The recommendation loss uses original candidates. The KL term is averaged over examples with at least one token-distinct replacement across the two views; unchanged views and minibatches without replacements contribute zero. All three distributions receive gradients. Auxiliary dropout is disabled, while supervised forwards retain dropout. Veracity-attribution pseudo-labels and their prediction branch are omitted; inference uses the ordinary recommender.

\section{Attack Generation and Training}
\label{app:attack_details}
\label{app:atr_details}
\label{app:prompt_details}

\paragraph{Defense-pool model checkpoints.}
The two fine-tuning bases are the official Mistral Small 3.2 24B Instruct (2506 release)~\citep{mistralsmall32_2025} and Olmo 3 7B Instruct~\citep{olmo3_2025} checkpoints. Each uses its own tokenizer, native instruction template, and end-of-turn markers. The Mistral model also generates the Source Bias and SheepDog-adapted training rewrites. Gemma 4 models are used only for evaluation attacks.

\paragraph{Prompt and decoding.}
The prompt-only baseline and LANCE Reflector share this instruction:
\begin{quote}
\small
You are an expert in news headline optimization for recommender systems. Given a news title and its category, rewrite the title to improve user engagement and recommendation relevance. Preserve the meaning, entities, events, quantities, attribution, and certainty expressed in the original title. Do not add unsupported facts. Use the category as context and keep it unchanged.

Write a headline of at most 30 words.

Output only this JSON object, with no extra keys or explanation:

\texttt{\{"title": "REWRITTEN TITLE"\}}

Title: \emph{[original title]}\\*
Category: \emph{[original category]}\\*
Output:
\end{quote}
Decoding uses temperature 0.95, top-$p$ 0.7, top-$k$ 50, at most 768 new tokens, and one response per target, with thinking disabled for the prompt-only Gemma 4 31B. A JSON schema constrains output; only nonempty titles extracted from complete responses are accepted. Structural failures retain the original title under Section~\ref{sec:metrics}. Generation uses the prompt's 30-word limit, and the recommender subsequently truncates titles to 30 tokens.

\paragraph{LANCE}
All generators use native instruction-tuned chat templates. Explorer preference pairs use exploration news, training user contexts, and a top-50 recommendation threshold. Each Reflector uses standard DPO~\citep{dpo2023} with its frozen initial instruction-tuned checkpoint as reference: Mistral Small 3.2 24B Instruct or Olmo 3 7B Instruct for defense-pool generation, and Gemma 4 E4B for evaluation attacks. LoRA leaves base weights frozen, with rank 8, scaling factor 16, and zero dropout. Loss covers the title response and end-of-turn token, excluding prompt tokens. Across recommendation backbones and Reflector base models, DPO uses learning rate $10^{-4}$, $\beta=0.1$, and three epochs, retaining the final checkpoint. One pair per microbatch and 24 accumulation steps give an effective batch size of 24. Other settings are a 1,536-token prompt-plus-response limit, cosine learning-rate decay, 0.1 warm-up ratio, gradient clipping at 1.0, and BF16 precision.

\paragraph{ATR-2FT adaptation.}
Following ATR~\citep{atr2024} and LANCE's news adaptation~\citep{lance2025}, a mapping $A_\phi$ aligns language-model representations $g_\psi(x)$ with the attacked recommender's frozen news encoder $e_\theta(x)$ on exploration news $\mathcal{V}_e$:
\begin{equation}
 \mathcal{L}_{\mathrm{align}}=\E_{x\in\mathcal{V}_e}
 \|A_\phi(g_\psi(x))-e_\theta(x)\|_2^2.
\end{equation}
After domain adaptation by next-token prediction of titles conditioned on categories, the generator minimizes
\begin{align}
 \mathcal{L}_{\mathrm{ATR}}&=\mathcal{L}_{\mathrm{gen}}
 +\lambda_{\mathrm{prom}}\mathcal{L}_{\mathrm{prom}},\\
 \mathcal{L}_{\mathrm{prom}}&=\sum_{u\in\mathcal{U}'}\sum_{n\in\mathcal{V}'_e}
 \max\{s^{\max}_u-\widehat{s}_\psi(u,n)+\gamma,0\}.
\end{align}
Here $\mathcal{U}'$ and $\mathcal{V}'_e$ are sampled training contexts and exploration news; $\widehat{s}_\psi$ uses the aligned representation, $\gamma$ is the margin, and $\lambda_{\mathrm{prom}}$ weights promotion. The maximum training-catalog score $s^{\max}_u$ is refreshed each epoch. The attacked recommender stays frozen: RQ1--RQ4 use the undefended checkpoint, whereas RQ5 uses each evaluated checkpoint.

\section{Evaluation Details}
\label{app:evaluation_details}

Catalog ranking excludes padding and breaks score ties by original catalog order. AUC uses average ranks for ties, nDCG averages gains within tied-score blocks, and MRR uses stable candidate order. User contexts are identified by user ID and complete original history before truncation; attacked histories are updated before recomputing user representations. Unchanged target ranks count as BSR failures. For Table~\ref{tab:recommendation_performance}, let $B_r$ and $A_r$ denote the unrounded Before and After scores for run $r$, respectively. We report the relative change as $\frac{1}{5}\sum_{r=1}^{5}100\times(A_r/B_r-1)\%$.

\end{document}